\documentclass[useAMS,usenatbib]{mn2e}

\usepackage{graphicx}
\usepackage{epstopdf}
\usepackage{amssymb}
\title[Bimodal morphologies of massive galaxies at the core of a protocluster at $z=3.09$]{
Bimodal morphologies of massive galaxies at the core of a protocluster at $\bf z=3.09$ and the strong size growth of a brightest cluster galaxy}
\author[Mariko Kubo et al.]{M. Kubo$^{1,2}$\thanks{E-mail:
mariko.kubo@nao.ac.jp},  
T.~Yamada,$^{3,4}$ T.~Ichikawa,$^{4}$ M.~Kajisawa,$^{5}$ Y.~Matsuda,$^{6,7}$ 
\newauthor
I.~Tanaka,$^{8}$ H.~Umehata$^{9,10}$\\
$^{1}$National Astronomical Observatory of Japan, Osawa 2-21-1, Mitaka, Tokyo 181-8588, Japan \\
$^{2}$Institute for Cosmic Ray Research, University of Tokyo, 5-1-5 Kashiwa-no-Ha, Kashiwa City, Chiba 277-8582, Japan\\
$^{3}$ISAS/JAXA, 3-1-1 Yunodai, Chuo-ku, Sagamihara, Kanagawa 252-5210,  Japan \\
$^{4}$Astronomical Institute, Tohoku University, 6-3 Aoba, Aramaki, Aoba-ku, Sendai, Miyagi 980-8578, Japan \\
$^{5}$Research Centre for Space and Cosmic Evolution, Ehime University, Bunkyo-cho 2-5, Matsuyama 790-8577,  Japan  \\
$^{6}$Chile Observatory, National Astronomical Observatory of Japan, Tokyo 181-8588, Japan \\
$^{7}$SOKENDAI (Graduate University for Advanced Studies), Osawa 2-21-1, Mitaka, Tokyo 181-8588, Japan \\
$^{8}$Subaru Telescope, National Astronomical Observatory of Japan, 650 North A'ohoku Place, Hilo, HI 96720, USA \\
$^{9}$Institute of Astronomy, The University of Tokyo, Mitaka, Tokyo 181-0015, Japan \\
$^{10}$The Open University of Japan, 2-11 Wakaba, Mihama-ku, Chiba 261-8586, Japan \\
}
\begin{document}

\date{in original form 2014 October 31}

\pagerange{\pageref{firstpage}--\pageref{lastpage}} \pubyear{}

\maketitle

\label{firstpage}

\begin{abstract}
We present the near-infrared high resolution imaging 
of an extremely dense group of galaxies
at the core of the protocluster at $z=3.09$ in the SSA22 field 
by using the adaptive optics AO188 and the Infrared Camera 
and Spectrograph (IRCS) on Subaru Telescope. 
Wide morphological variety of them suggests their on-going dramatic evolutions.  
One of the two quiescent galaxies (QGs), the most massive one in the group, 
is a compact elliptical with an effective radius $r_{e} = 1.37\pm0.75$ kpc. 
It supports the two-phase formation scenario of giant ellipticals today
that a massive compact elliptical is formed at once and evolves 
in the size and stellar mass by series of mergers. 
Since this object is a plausible progenitor of a brightest cluster galaxy (BCG) 
of one of the most massive clusters today,
it requires strong size ($\ga10$) and stellar mass ($\sim$ four times by $z=0$) growths. 
Another QG hosts an AGN(s) and is fitted with a model composed 
from an nuclear component and S\'ersic model. 
It shows spatially extended [O{\footnotesize III}]$\lambda$5007 emission line  
compared to the continuum emission, a plausible evidence of outflows. 
Massive star forming galaxies (SFGs) 
in the group are two to three times larger than the field SFGs at similar redshift. 
Although we obtained the $K$-band image deeper than the previous one, 
we found no candidate new members. 
This implies a physical deficiency of low mass galaxies 
with stellar mass $M_{\star}\la4\times10^{10}~M_{\odot}$ 
and/or poor detection completeness of them owing to their diffuse morphologies.
\end{abstract}

\begin{keywords}
galaxies: formation  --- galaxies: evolution --- galaxies: distances and redshifts  --- galaxies: clusters: general
\end{keywords}

\section{Introduction}
There is the well established morphology-density relation in the current Universe 
where elliptical and S0 galaxies dominate rich cluster cores 
while spiral galaxies are dominant in general fields \citep{1980ApJ...236..351D}. 
The galaxy morphologies are tightly related to properties of galaxies that 
massive early-type galaxies (ETGs) are generally dominated by old stars,  
and with low star formation activities and gas contents. 
The physical mechanisms that relate morphological transformation
and shutting down star formation activity 
with the environment are still open questions. 
In mature clusters in the current Universe, 
harassment \citep{1996Natur.379..613M}, strangulation \citep{1980ApJ...237..692L} 
and ram-pressure stripping \citep{1972ApJ...176....1G}
can play important roles on quenching star formation 
and transforming morphologies, but on the other hand, 
red sequences have already appeared in the protoclusters 
at $z=2-3$ \citep{2007MNRAS.377.1717K, 2008ApJ...680..224Z, 2008ASPC..399..373U,2012ApJ...750..116U,2013ApJ...778..170K}, 
when the galaxy clusters have not yet been fully virialized. 

Massive quiescent galaxies (QGs) at up to $z\sim3$ are now found by deep multi-wavelength surveys in general fields (e.g., \citealt{2013A&A...556A..55I,2013ApJ...777...18M,2016ApJ...820...11M}).  
Generally, they are remarkably compact compared to massive ETGs today
(e.g., \citealt{2005ApJ...626..680D, 2006ApJ...650...18T, 2007ApJ...671..285T, 2008ApJ...677L...5V, 2009ApJ...695..101D, 2010ApJ...709.1018V, 2014ApJ...788...28V}).  
\citet{2010ApJ...709.1018V} and \citet{2013ApJ...778..115P}
argue that such compact QGs at $z>2$
are plausible progenitors of massive ETGs 
by comparing constant cumulative number density samples 
of massive galaxies from $z=0$ to $3$. 
The dissipative processes like gas rich major mergers \citep{2006ApJ...650..791C,2007ApJ...658..710N, 2010ApJ...722.1666W,2011ApJ...730....4B} and/or in-streaming gas by violent disk instabilities \citep{2009ApJ...703..785D, 2015MNRAS.447.3291C} are proposed as formation scenarios of such compact elliptical galaxies. 
Series of dry minor mergers can increase the sizes 
of compact QGs effectively (e.g., \citealt{2009ApJ...699L.178N}). 
Based on high resolution cosmological numerical simulations, 
\citet{2010ApJ...725.2312O} proposed the two phase formation scenario 
in which in-situ rapid gas accretion and violent star formation 
form a compact spheroid at first and it is grown by mergers of galaxies  
formed at outside of its virial radius. 

One of the major uncertainties in the previous studies 
is the traceability of progenitors of massive ETGs. 
Protoclusters are suitable targets to study progenitors of galaxies dominating rich cluster cores today. 
The strong size growths of massive ETGs is supported from many studies of QGs 
in the protoclusters at $z<2$ \citep{2012ApJ...744..181Z, 2012MNRAS.419.3018C, 2012ApJ...750...93P, 2013ApJ...773..154L,2014ApJ...788...51N, 2016A&A...593A...2A}. 
But it is not still unclear how their evolutions relate to environments at earlier time. 
To challenge this question, we need to study protoclusters at the epoch
when the morphology-density relation just arises. 

The SSA22 protocluster at $z=3.09$ is a rare density peak of galaxies 
found from the overdensity of Lyman break galaxies (LBGs; \citealt{1998ApJ...492..428S})
at first and in later well characterized as a core high density region of a superstructure 
by the wide field (1.38 deg$^2$ in the SSA22) narrow-band surveys 
of Ly$\alpha$ emitters (LAEs) at $z\approx3.09$ \citep{2004AJ....128.2073H, 2012AJ....143...79Y}. 
The velocity dispersion of the protocluster core is $\sim700$ km s$^{-1}$  
and the cluster mass measured by using the velocity dispersion 
or overdensity are $\sim2-5\times10^{14}~M_{\odot}$ \citep{2015ApJ...799...38K}. 
Thus this protocluster is a plausible progenitor of a core of one of the most massive clusters today. 
In \citet{2013ApJ...778..170K}, we reported that there is an overdensity 
of massive galaxies ranging from active SFGs to massive QGs in the SSA22 protocluster. 
This suggests that a red sequence has just begun appearing in this protocluster. 

\citet{2012ApJ...750..116U} discovered dense groups of massive galaxies 
as the counterparts of Ly$\alpha$ Blobs (LABs) 
and sub-mm galaxies (SMGs) in the SSA22 protocluster. 
Some of them are spectroscopically confirmed 
as plausibly physically associated groups in \citet{2016MNRAS.455.3333K}. 
Similarly, by the abundance matching technique in a wide field survey, 
\citet{2016ApJ...816...86V} reported massive galaxies surrounded 
by many companions at high redshift as the progenitors of ultra massive galaxies today.
They are likely to be hierarchical multiple mergers at the early-phases 
of the formation histories of massive ETGs, predicted from
the high resolution cosmological numerical simulations in the $\Lambda$CDM Universe
(e.g., \citealt{2003ApJ...590..619M, 2007ApJ...658..710N})
and thus excellent laboratories of morphological evolutions of massive ETGs. 

We here present the deep and high resolution imaging of 
an extremely dense group of galaxies at the core of the SSA22 protocluster, 
called the SSA22-AzTEC14 group, in the $K'$-band 
by using the InfraRed Camera and Spectrograph (IRCS; \citealt{2000SPIE.4008.1056K})
and the Adaptive Optics system AO188 \citep{2010SPIE.7736E..0NH} on Subaru Telescope. 
Since the {\it F160W}-band of Hubble Space Telescope ({\it HST}), 
is too blue to study stellar morphologies in the rest-frame optical of galaxies at $z>3$, 
an AO assisted $K$-band (rest-frame $V$-band) imaging with a 10-m class ground-based telescope is the best option for our targets. 
We describe the observation in Section 2. 
In Section 3, we report morphologies of galaxies in the AzTEC14 group.   
We discuss the environmental dependence of morphologies  
of massive QGs and SFGs, and behaviors of faint galaxies in the group in Section 4.  
In this paper, cosmological parameters 
of $H_0 = 70$ km s$^{-1}$ Mpc$^{-1}$ , $\Omega_{\rm m} =0.3$ 
and $\Omega_{\Lambda} = 0.7$ are assumed.  
In this cosmology, 1 arcsec corresponds to $7.633$ kpc in physical at $z=3.09$.
We adopt the \citet{2003PASP..115..763C} Initial Mass function (IMF). 
The AB magnitude system is used throughout this paper. 

\begin{figure} 
\begin{center}
\includegraphics[width=80mm]{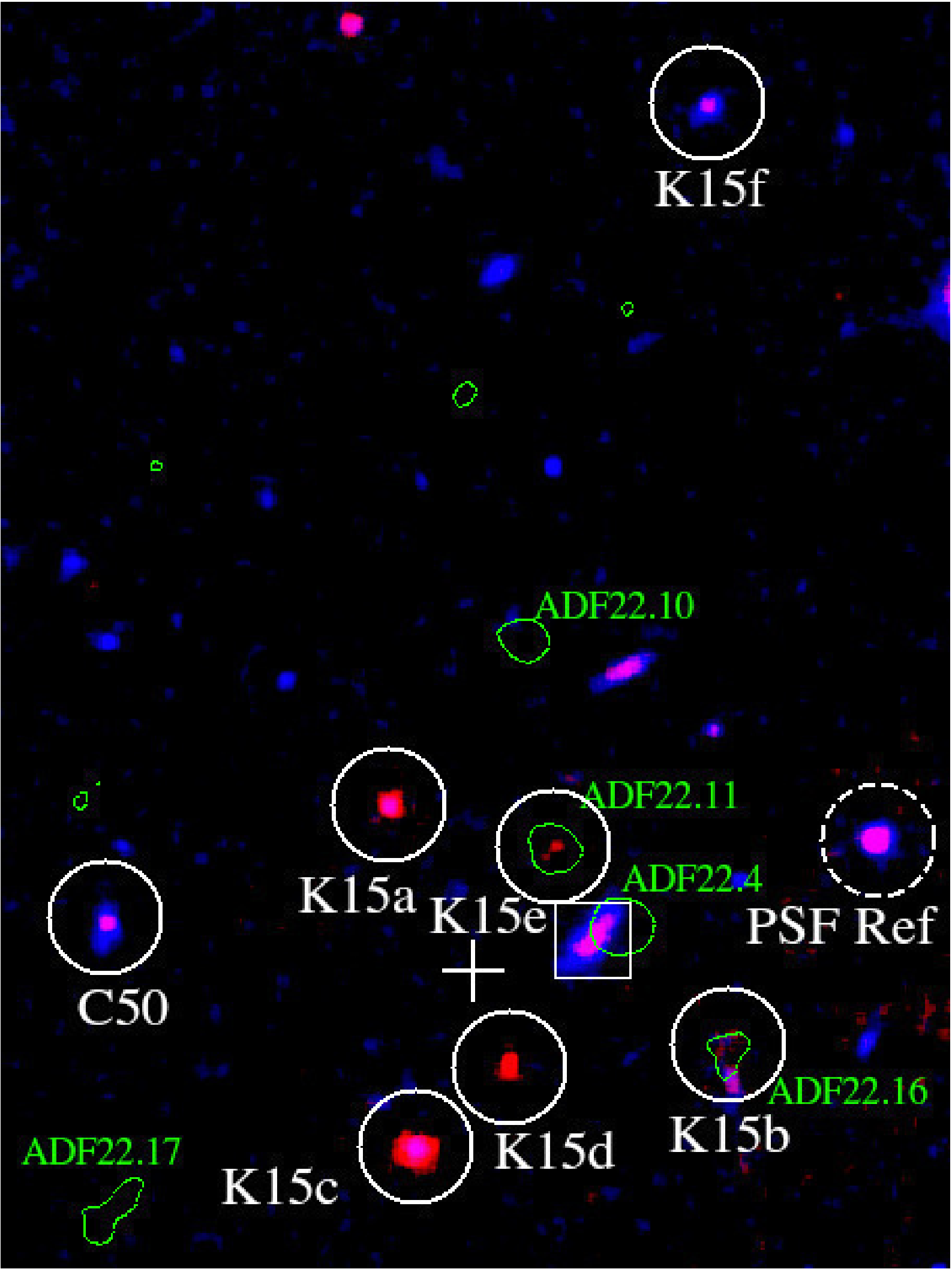}
\caption{
The combined image of the IRCS-AO $K'$ (red) and {\it HST/ACS} $I_{\rm F814W}$ (blue)-band images 
of the AzTEC14 group ($20''.0\times27''.0$).  
The white circles show the objects 
with spectroscopic redshifts $z_{\rm spec}\approx3.09$. 
The object IDs are the same as those in \citet{2016MNRAS.455.3333K} and \citet{2017ApJ...835...98U}. 
The white cross shows the coordinate of the LGS set in our operation. 
The dashed white circle shows the PSF reference star adopted in this study. 
The green contours indicate the 1.1 mm sources detected above $3\sigma$ per pixel 
on the image corrected with ALMA by \citet{2017ApJ...835...98U}. 
The white square shows the spectroscopically confirmed outlier. 
} 
\label{fig:az14ao-all}
\end{center}
\end{figure}

\section[Observation]{Observation}

Our target is an extremely dense group of galaxies found at the core 
of the SSA22 protocluster at $z=3.09$, called the SSA22-AzTEC14 group \citep{2016MNRAS.455.3333K}. 
Fig. \ref{fig:az14ao-all} shows its combined image 
of the IRCS-AO $K'$ (red) and {\it HST F814W} 
(blue, here after {\it HST} $I_{\rm F814W}$)-band images. 
The object IDs are the same as those in \citet{2016MNRAS.455.3333K} and \citet{2017ApJ...835...98U}.
The AzTEC14 group was first discovered as a rare overdensity of distant red galaxies (DRGs; $J-K>1.4$)
at the position of a bright 1.1 mm source found from the ASTE/AzTEC 1.1 mm survey of the SSA22 field  
\citep{2009Natur.459...61T,2014MNRAS.440.3462U} by \citet{2012ApJ...750..116U}.  
In \citet{2016MNRAS.455.3333K}, we spectroscopically confirmed that
seven galaxies belong to one group at $z_{\rm spec}\approx3.09$.
Note that there is a large redshift uncertainty for Az14-K15c 
as its redshift was measured with the Balmer / 4000 \AA~ breaks 
of its continuum spectrum. 
Five of them have stellar masses $M_{\star}>10^{10.5}~M_{\odot}$ 
and five of them are classified as DRGs. 
Comparing the AzTEC14 group with the galaxy formation models based 
on the Millennium simulation \citep{2005Natur.435..629S}, 
we found that this group has properties similar to those 
of a dense group of galaxies at high redshift 
which evolves into a brightest cluster galaxy (BCG) 
of one of the most massive clusters in the current Universe \citep{2016MNRAS.455.3333K}. 
Moreover, we carried out deep observations of this region at sub-mm 
by using Atacama Large Millimeter/submillimeter Array (ALMA)  
with a synthesized beam of $0''.70\times0''.59$ and a typical rms level of 60 $\mu$Jy beam$^{-1}$ 
(green contours in Fig. \ref{fig:az14ao-all}, \citealt{2015ApJ...815L...8U}; \citealt{2017ApJ...835...98U}). 
The 1.1 mm fluxes of five sub-mm sources detected in the AzTEC14 group 
range from $S_{\rm 1.1~mm} = 0.6$ to $2.0$ mJy. 
ADF22.4 in Fig. \ref{fig:az14ao-all} is newly confirmed at $z_{\rm spec}=3.091$ 
by detecting the redshifted CO(9-8) emission line \citep{2017ApJ...835...98U} and [C {\footnotesize II}] 158 $\mu$m (Hayatsu et al. submitted). 
Then now eight galaxies are confirmed as a dense group of galaxies at $z_{\rm spec}\approx3.09$. 

Our high resolution near-infrared (NIR) imaging observation of the AzTEC14 group 
was conducted on 24, July 2015 by using the IRCS and AO188 equipped 
on Subaru Telescope (S15A-059; PI Mariko Kubo). 
The IRCS was used in the 52 mas plate scale mode with 54 arcsec field of view, 
and with the $K'$-band filter. 
The AO188 was operated in the laser guide star AO (LGSAO) mode. 
The tip tilt guide star (TTGS) for the LGSAO operation is 
a star with $R=18.0$ at (R.A., Dec) = (22:17:35.78 +00:19:16.3) 
which is $52-64$ arcsec apart from the targets. 
The exposure time was 2.8 hours in total. 
We reduced the data using the {\sf IRAF} data reduction tasks 
following the data reduction manual for the IRCS\footnote{http://www.subarutelescope.org/Observing/DataReduction/C\\ookbooks/IRCSimg\_2010jan05.pdf}.
The individual frames were combined after masking the bad pixels, flat fielding, sky subtraction 
and estimating the dither offsets in a standard manner. 
The zero-point magnitude of our IRCS-AO $K'$-band image
is calibrated to that of our MOIRCS $K_s$-band image.

Thanks to the good observing condition, 
the AO works well despite the use of a faint and distant TTGS.
The FWHM of the Point Spread Function (PSF) size 
at the PSF reference star　(the dashed white circle in Fig. \ref{fig:az14ao-all}) is $\approx0''.16$ 
while that in our previous MOIRCS imaging at this field is $\approx0''.41$. 
The 5$\sigma$ limiting magnitude measured 
with $0''.3$ and $1''.1$ diameter apertures on 
our IRCS-AO $K'$-band image are $K=25.49$ and $23.51$, respectively 
while that measured with an $1''.1$ diameter aperture 
on our MOIRCS $K_s$-band image is $K=24.3$.
Thus, for galaxies extended over $1''.1$ ($\approx 8$ kpc), 
detection completeness on our IRCS-AO $K'$-band image 
is lower than that on our MOIRCS $K_s$-band image. 

We also show the archival $I_{\rm F814W}$-band image taken 
with the Advanced Camera for Surveys (ACS) on the {\it HST} (PID 9760; PI Roberto Abraham). 
The FWHM of the PSF size on the $I_{\rm F814W}$-band image is $0''.08$ 
and the 5$\sigma$ limiting magnitude measured 
with a $0''.3$ diameter aperture is $I_{\rm F814W}=28.3$. 

\begin{figure*} 
\begin{center}
\includegraphics[width=175mm]{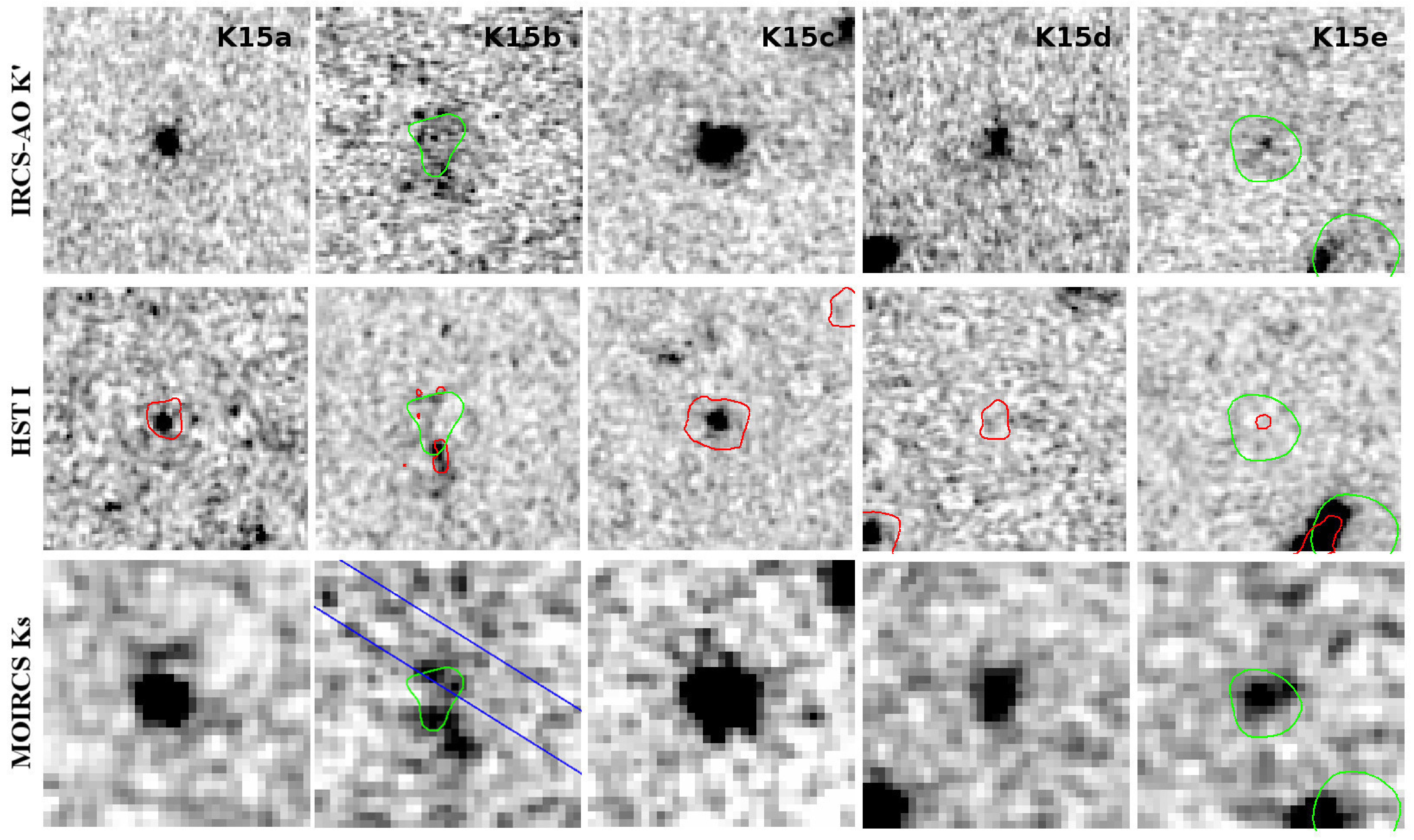}
\includegraphics[width=175mm]{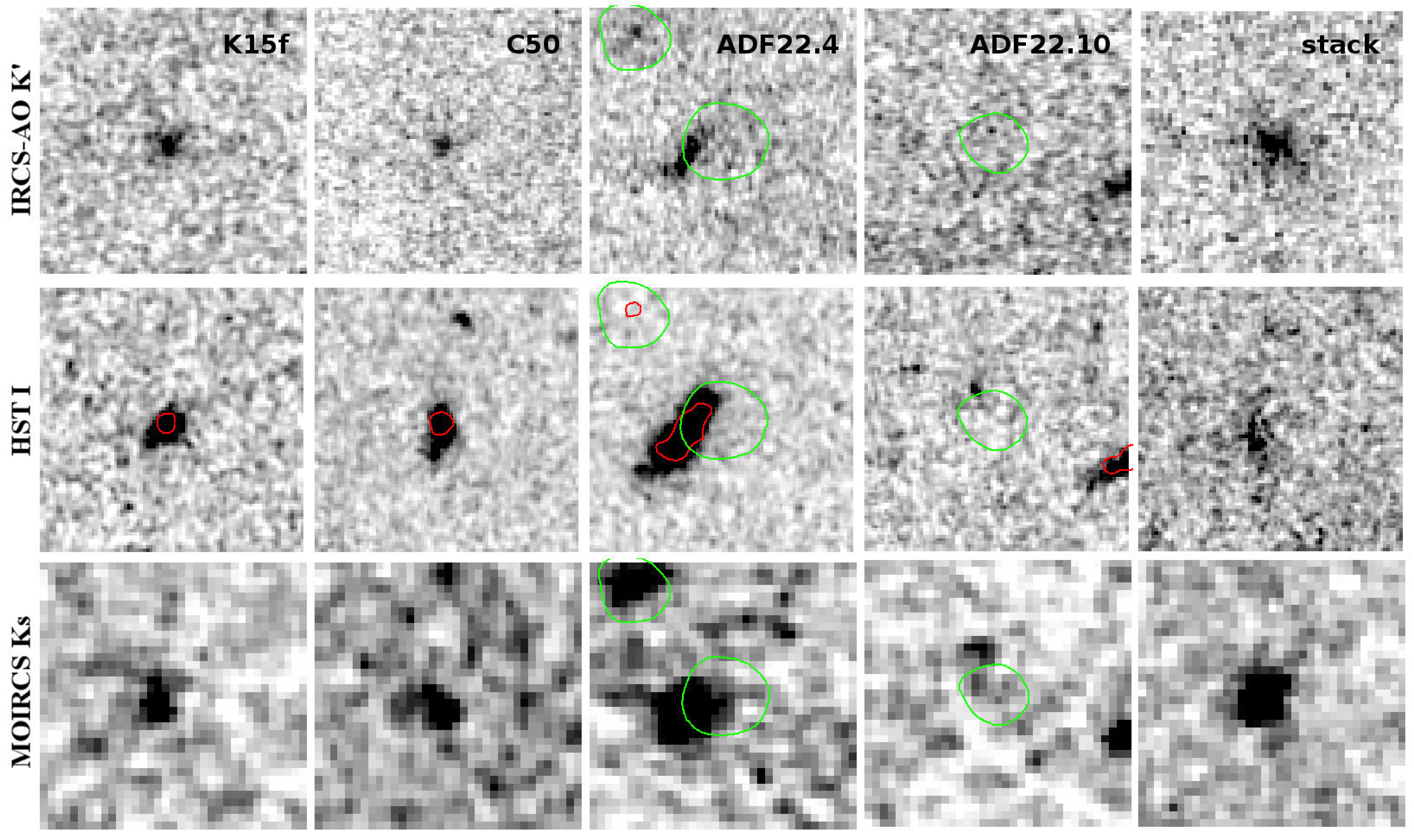}
\caption{
The {\it top}, {\it middle} and {\it bottom} raws show the IRCS-AO $K'$, 
{\it HST}/ACS $I_{\rm F814W}$ and MOIRCS $K_s$-band images 
of members of the AzTEC14 group, respectively ($4''.1\times4''.1$). 
The object IDs are the same as those in Fig. \ref{fig:az14ao-all}. 
We show the stacked images of SFGs (Az14-K15b, K15d, K15e, K15f \& C50) at the right-bottom-end panels. 
The red contours shown in the {\it middle} raws are isophotal areas 
of the sources detected on the IRCS-AO $K'$-band image above $2\sigma$ per pix. 
The green contours show isophotal areas of the 1.1 mm sources 
detected with ALMA above $3\sigma$ \citep{2017ApJ...835...98U}. 
The blue solid lines drawn on the MOIRCS $K_s$-band image of Az14-K15b
show the slit position of our MOIRCS spectroscopy 
of this object \citep{2015ApJ...799...38K}. 
} 
\label{fig:az14ao-image}
\end{center}
\end{figure*}

\begin{table*}
 \centering
  \caption{{\sf GALFIT} Morphological Parameters}
\begin{tabular}{lccccccc}
\hline \hline
Object  & $K_{\rm tot}$ & $M_{\ast}$ & $r_e$ & S\'ersic $n$ & b/a & PA & $\chi^2$/dof \\
              & (mag)  & ($10^{10}~M_{\odot}$) & (kpc) & & & (deg) & \\
\hline
Az14-K15a & $22.50\pm0.10$  & $ 8.0_{- 2.9 }^{+ 5.7 } $ &  $1.37\pm0.75$ & $9.5\pm4.5$ & $0.58\pm0.09$ & $-61\pm8$ & 1.18\\
Az14-K15a (AGN-subtracted) &     &                   &  $4.26\pm0.60$ & $2.2\pm0.6$ & $0.60\pm0.09$ & $-62\pm12$ & 1.18\\
Az14-K15c & $21.55\pm0.04$  & $ 25.4_{- 5.8 }^{+ 7.3} $  &  $1.01\pm0.04$ & $2.5\pm0.2$ & $0.89\pm0.03$ & $-5\pm11$ & 1.10\\
Az14-K15d & $23.06\pm0.16$  & $ 5.3_{- 2.9 }^{+ 5.7 } $  &  $3.23\pm1.56$ & $4.6\pm1.8$ & $0.35\pm0.07$ & $85\pm4$ & 1.06\\
Az14-K15e & $23.35\pm0.21$  & $ 7.2_{- 5.2 }^{+9.8} $  &  $11.9\pm27.1$ & $7.3\pm8.9$ & $0.09\pm0.05$ & $35\pm3$ & 1.07\\
Az14-K15f  & $23.30\pm0.20$  & $ 1.7_{- 0.8 }^{+ 2.2 } $   &  $1.76\pm0.32$ & $1.3\pm0.4$ & $0.90\pm0.12$ & $-10\pm52$ & 0.91\\
C50              &  $24.09\pm0.37$ & $ 1.1_{- 0.7 }^{+ 1.1 } $   &  $1.84\pm1.24$ & $3.9\pm2.8$ & $0.64\pm0.19$ & $78\pm20$ & 0.99\\
\hline
\end{tabular}
\label{tab:tableobjects1}
\end{table*}

\section[Results]{Results}
Figure \ref{fig:az14ao-image} shows the IRCS-AO $K'$, 
{\it HST}/ACS $I_{\rm F814W}$ and MOIRCS $K_s$-band images of the galaxies in the AzTEC14 group. 
We also show the images at the position of ADF22.10 
though it has not yet been spectroscopically confirmed as a group member. 
The red contours on the $I_{\rm F814W}$-band stamps show 
the regions detected above $2\sigma$ per pixel on the IRCS-AO $K'$-band image.  
The green contours show isophotal areas of the 1.1 mm sources same as Fig. \ref{fig:az14ao-all}.
The bottom right end raws show the stacks of the Az14-K15b, d, e, f and C50, 
the members classified as SFGs in \S~3. 1. 
We mistook the object ID C50 as MD048 in the previous paper. 
Before stacking the images, the image centres are aligned 
at the centroids of the objects on the IRCS-AO $K'$-band image 
or MOIRCS $K_s$-band image if they are not detected on our IRCS-AO $K'$-band image.  
Both the $K$-band images are combined in median after matching their scales 
based on their total magnitudes measured on our MOIRCS $K_s$-band images.  
The $I_{\rm F814W}$-band images are combined without any scaling 
since many of the group members are hard to be identified on the $I_{\rm F814W}$-band images individually. 

It is interesting that a wide variety of galaxies are observed within such a small volume. 
Even if we focus on only the galaxies with $M_{\star}\sim10^{11}~M_{\odot}$, 
they have a wide variety in morphologies, 
similar to a dense cluster reported in \citet{2016ApJ...828...56W} recently. 
Such extremely dense groups at high redshift are interesting 
laboratories maybe just at the transition epoch of morphologies.

Many of the members are hardly detected on the $I_{\rm F814W}$-band image 
maybe owing to their red colors. 
Therefore the AO-assisted high-resolution $K$-band imaging is essential to study
morphologies of such red galaxies at $z>3$. 
Interestingly, some galaxies are more clearly detected on our MOIRCS $K_s$-band image 
(K15b and K15e in Fig. \ref{fig:az14ao-image})
while the point source sensitivity of our IRCS-AO $K'$-band image 
is better than that of our MOIRCS $K_s$-band image. 
We test the dependence of detection completeness 
on source morphology in \S~3. 3. 3. 

\subsection[Morphologies and stellar populations]{Morphologies and stellar populations}

\begin{figure}
\begin{center}
\includegraphics[width=75mm]{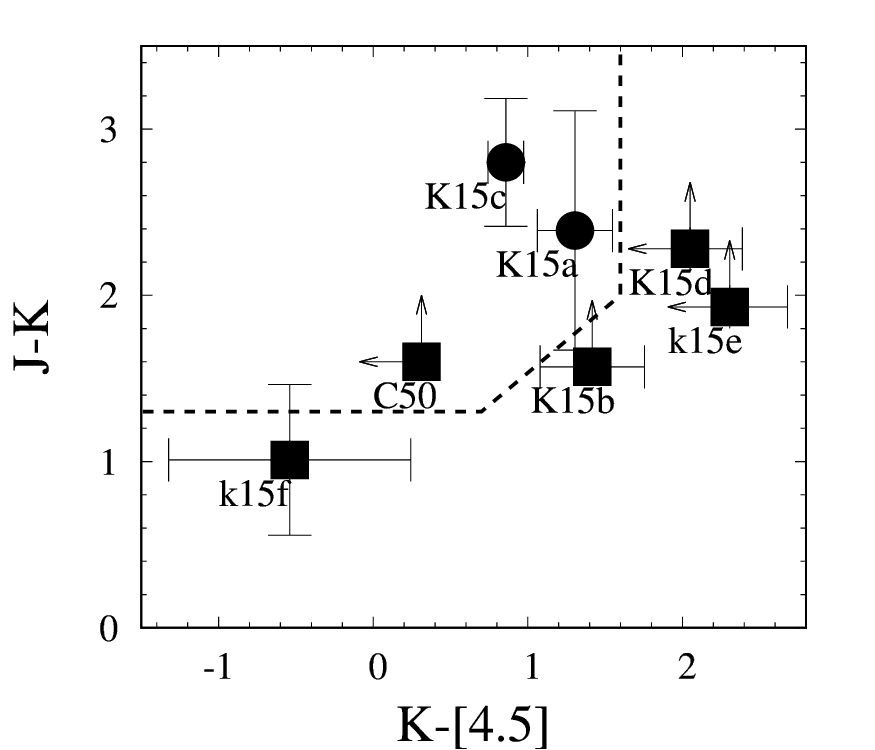}
\caption{
$J-K$ v.s. $K-[4.5]$ or rest-frame $UVJ$ color diagram of galaxies in the AzTEC14 group. 
The black dashed line shows the rest-frame $UVJ$ color criterion for QGs in previous studies. 
The black filled circles and squares show the galaxies classified as QGs 
and SFGs in the AzTEC14 group, respectively.
C50 looks satisfying QG color criterion but there is a large uncertainty in its rest-frame $UVJ$ color.
K15d and K15e suffer from deblendings with adjacent sources on the 4.5 $\mu$m-band image. 
} 
\label{fig:az14-uvj}
\end{center} 
\end{figure}

At first, we classify the members into QGs and SFGs 
from their rest-frame $UVJ$ colors and SEDs. 
Fig. \ref{fig:az14-uvj} shows $J-K$ v.s. $K-[4.5]$ or 
rest-frame $UVJ$ color diagram \citep{2009ApJ...691.1879W}. 
Aperture corrected photometries of each galaxies 
are performed by the way same as that in \citet{2013ApJ...778..170K} but here we subtract 
spectroscopically measured H$\beta$ and [O{\scriptsize III}] $\lambda\lambda 5007$ 
emission line fluxes from their $K$-band fluxes. 
In \citet{2016MNRAS.455.3333K}, we showed their observed and best-fit model SEDs at 
the $u^{\star}, B, V, R, i', z', J, H, K$, 3.6, 4.5, 5.8 \& 8.0 $\mu$m-bands 
obtained by fitting the observed flux values to the stellar population 
synthesis models of the \citet{2003MNRAS.344.1000B} 
adopting the \citet{2003PASP..115..763C} Initial Mass Function. 

The two brightest members, Az14-K15a and K15c, 
satisfy the rest-frame $UVJ$ color criterion of QGs and 
also have SEDs well characterized as those of QGs \citep{2016MNRAS.455.3333K}. 
Other galaxies are classified as SFGs: 
C50 looks satisfying QG color criterion but there is large uncertainty in its rest-frame $UVJ$ color.
Looking its overall SED shown in \citet{2016MNRAS.455.3333K}, 
C50 has a blue color like a young SFG.  
Note that $K-[4.5]$ colors of Az14-K15d and K15e suffer from deblendings 
with adjacent sources at 4.5 $\mu$m. 
They can satisfy QG color criterion but are too faint 
to solve degeneracies of QG/SFG by SED fittings with our current data. 

\begin{figure} 
\includegraphics[width=85mm]{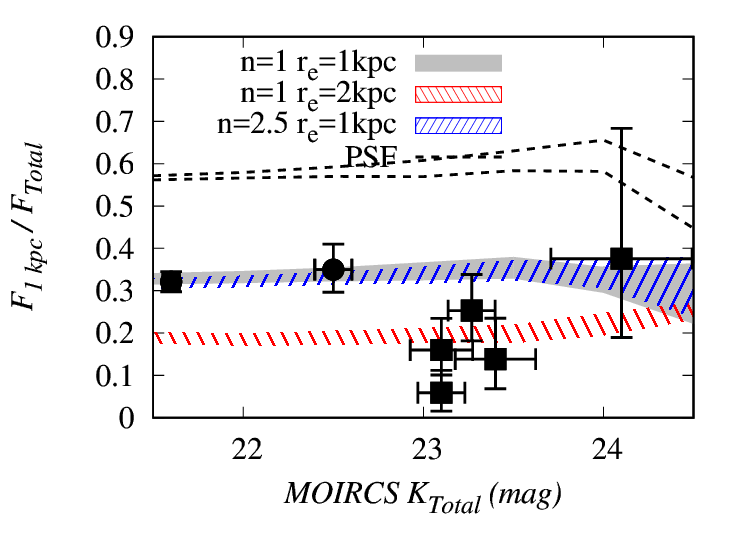}
\caption{
The ratios between central 1 kpc aperture radius 
fluxes measured on our IRCS-AO $K'$-band image
and total fluxes measured on our MOIRCS $K_s$-band image
as a function of total magnitudes measured on the MOIRCS $K_s$-band image. 
The galaxies classified as QGs and SFGs are shown with black filled circles and squares, respectively. 
The gray and red dashed regions show the S\'ersic models with $n=1$ and $r_e=1~\&~2$ kpc 
within 2 $\sigma$ rms ranges, respectively. 
Model flux ratios are properly calculated in the same manner 
as observed flux ratio estimated from both the $K$-band images. 
The blue dashed region is similar to gray dashed region but for a model with $n=2.5$ and $r_e=1$ kpc. 
The area between the black dashed lines shows a point source. 
} 
\label{fig:moircs-ao2}
 \end{figure}

Following previous studies, we evaluate morphological properties
by using the {\sf GALFIT} \citep{2002AJ....124..266P, 2010AJ....139.2097P}.  
The {\sf GALFIT} fits two-dimensional analytical functions
convolved with an PSF to observed galaxy images. 
We use a star with $K=22.6$ at (R.A., Dec) = (22:17:36.608, +00:18:22.52), 
which is 54, 9 and $5-16$ arcsec apart 
from the TTGS, LGS and targets, respectively, 
as a PSF reference star (the dashed white circle in Fig. \ref{fig:az14ao-all}). 
We fit the S\'ersic models \citep{1968adga.book.....S} 
with effective radii $r_e=0.2-12$ kpc and S\'ersic index $n=0.5-10$.  
Fittings are performed for within two arcsec square regions of each object. 
Sky background values are estimated at the areas 
$1.5$ to $2.0$ arcsec apart from each object before S\'ersic model fittings. 
We initially input total magnitudes measured by using the {\sf SExtractor} \citep{1996A&AS..117..393B} 
and typical morphological parameters for $z\sim3$ galaxies, $r_e = 1$ kpc and $n=1.4$. 
We summarize the morphological properties obtained 
by using the {\sf GALFIT} in Table \ref{tab:tableobjects1}. 

Since there are large uncertainties in 
the morphological parameters estimated with the {\sf GALFIT} 
except for those of the brightest one,  
we supplementary　compare observed central to total flux ratios 
with those of models in Fig. \ref{fig:moircs-ao2}. 
We take the 1 kpc radius aperture fluxes measured 
on the IRCS-AO $K'$-band image as central fluxes. 
Here we use the Kron fluxes measured on the MOIRCS $K_s$-band image 
by using the {\sf SExtractor} as total fluxes 
since up to 90\% of total fluxes of the galaxies 
in the AzTEC14 group measured on our MOIRCS $K_s$-band image  
are under the surface detection limit on our IRCS-AO $K'$-band image. 
We show the 2 $\sigma$ rms ranges of model flux ratios 
similarly measured on mock galaxy images in both the $K$-band 
from a thousand iteration for each. 
The two brightest members, Az14-K15a and Az14-K15c, 
have flux ratios similar to those of compact objects with $r_e\sim1$ kpc. 
On the other hand, except for the faintest one, 
SFGs have flux ratios lower than those of compact objects with $r_e\sim1$ kpc, 
i. e., more extended than typical SFGs at $z\sim3$. 
These results support the results of morphological analysis 
with the {\sf GALFIT} in the following. 

\begin{figure*}
\begin{center}
\includegraphics[width=170mm]{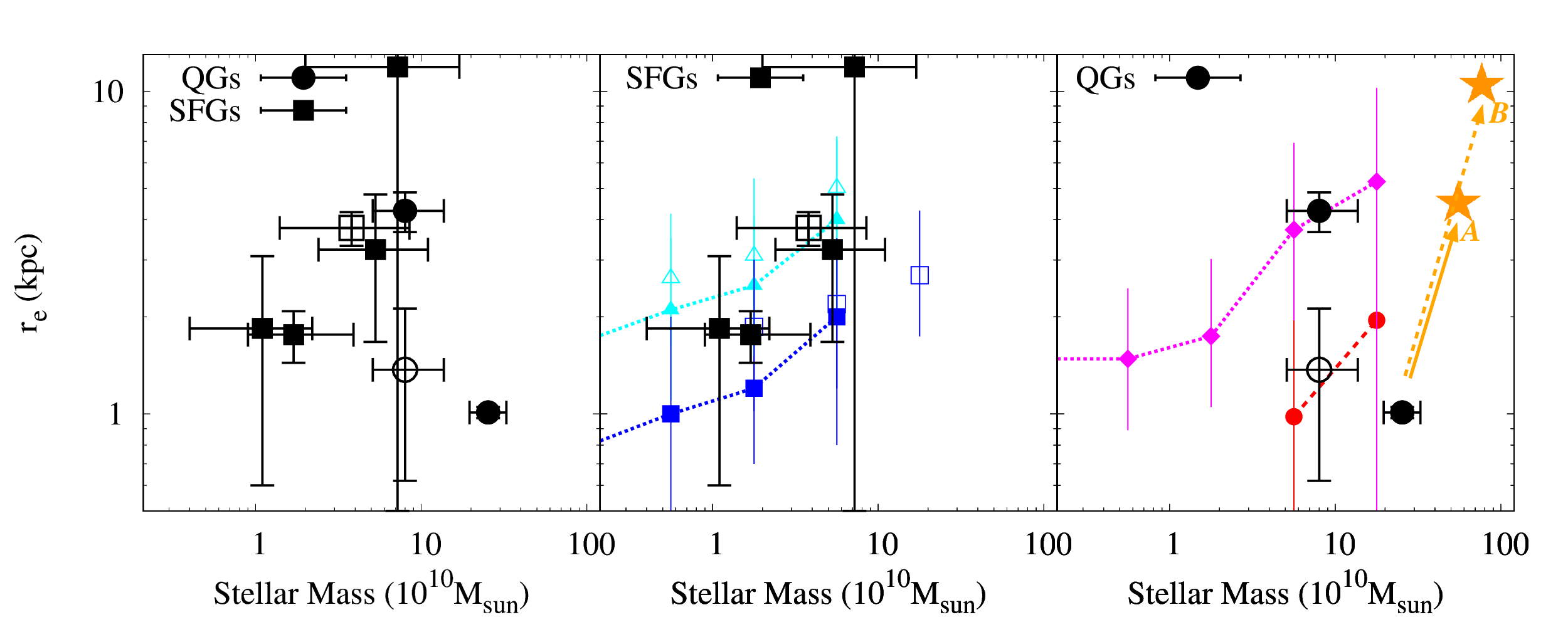}
\caption{
{\it Left:} The effective radius ($r_e$) to stellar mass relation. 
The black filled circles and squares show the QGs 
and SFGs in the AzTEC14 group, respectively. 
The large black open square shows the stack of the SFGs.
The large black open circle shows the result of a single S\'ersic fit of Az14-K15a. 
{\it Central:} Focusing on SFGs. 
The cyan filled triangles and small blue filled squares with dot lines show 
SFGs at $z=0$ and $z=3$ from S15, respectively. 
The cyan open triangles and small blue open squares show 
SFGs at $z=0.25$ and $z=2.75$ from vdW14, respectively. 
{\it Right:} Focusing on QGs. 
The magenta filled diamonds with dot line and 
small red filled circles with dot line show QGs $z=0$ and $z=3$ from vDW14, respectively. 
The orange stars at the point of the arrows are 
the expected sizes and stellar masses of Az14-K15c at $z\sim1$ 
in cases all the members merge into this object without (Case A) or with (Case B) in situ star formation. 
} 
\label{fig:az14-re}
\end{center} 
\end{figure*}

\begin{figure}
\begin{center}
\includegraphics[width=75mm]{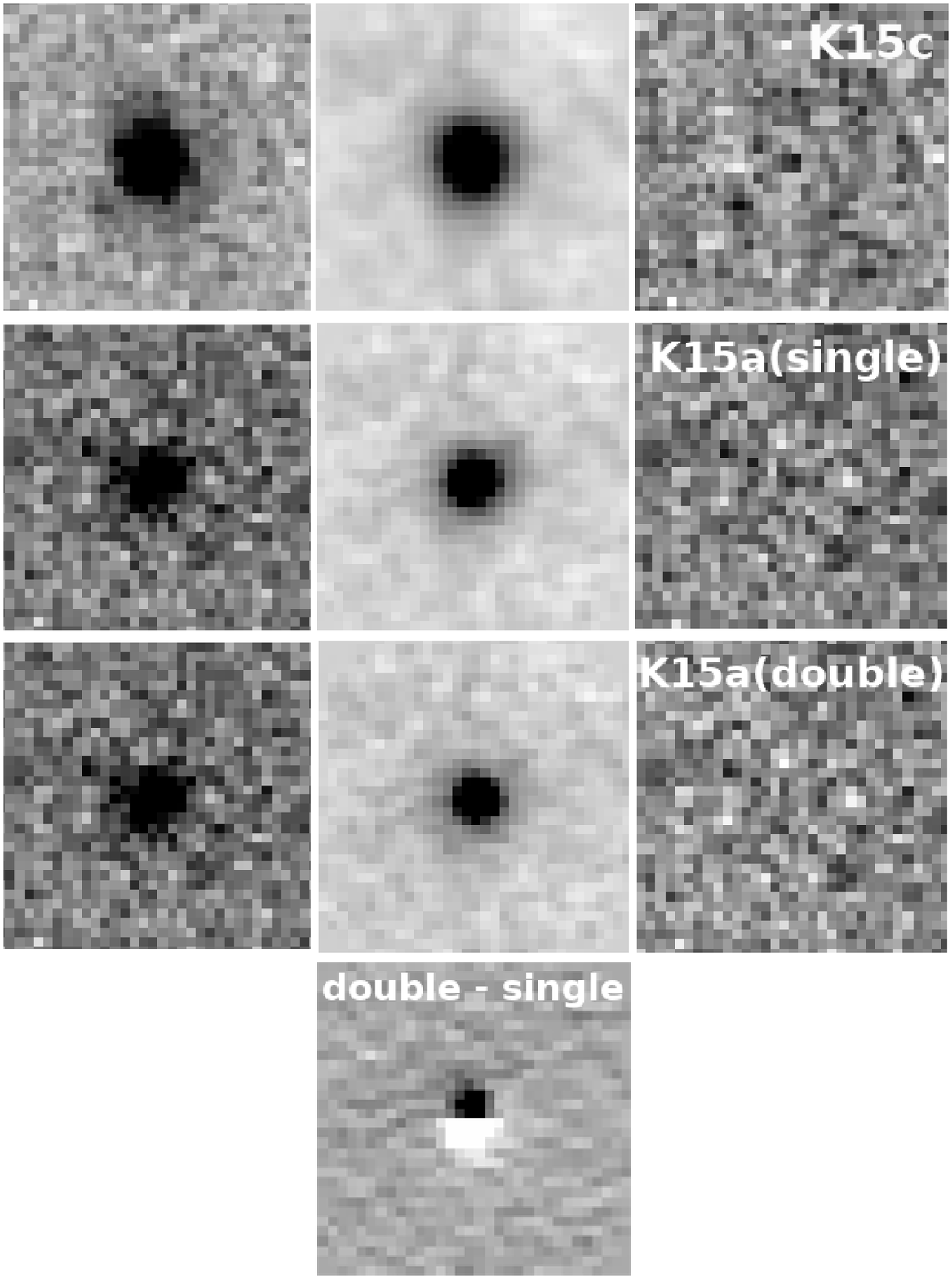}
\caption{
Examples of the S\'ersic model fittings of Az14-K15c ({\it first} row) and Az14-K15a 
({\it second} row: single S\'ersic model and {\it third} row: double-components model). 
The observed images, best-fit S\'ersic models and residuals from left to right. 
The sizes of the images are $1''.7$ square.
The {\it bottom} row shows the double-components model subtracted of single S\'ersic model. 
The scale is changed from the top three panels to emphasize differences. 
} 
\label{fig:residual}
\end{center} 
\end{figure}

\subsubsection[Quiescent galaxies]{Quiescent galaxies}

Figure \ref{fig:az14-re} shows the size-stellar mass distribution of galaxies in the AzTEC14 group. 
QGs are shown with black filled circles.  
We also plot the size-stellar mass relations 
of SFGs and QGs at $z\sim0$ and $z\sim3$ 
in vdW14 and S15, both obtained 
by using the Cosmic Assembly Near-infrared Deep Extragalactic Legacy Survey 
(CANDELS; \citealt{2011ApJS..197...35G,2011ApJS..197...36K}) 
and 3D-HST \citep{2012ApJS..200...13B} data. 
Comparison with other studies is done by consistent way; 
We compare the results in circularized effective radii and stellar masses measured adopting the Chabrier IMF; 
The galaxies in vdW14 and S15 are selected based on the spectroscopic 
and photometric redshifts, and classified by the rest-frame $UVJ$ color criterion; 
Morphological parameters in vdW14 and S15 are mainly evaluated  
by using the {\it HST F125W }\& {\it F160W}-band images.
In S15, SFGs at $z=3$ are studied by using the rest-frame UV data 
but they reported that morphological $k$-correction for them is less than $20\%$. 

\begin{figure*}
\begin{center}
\includegraphics[width=175mm]{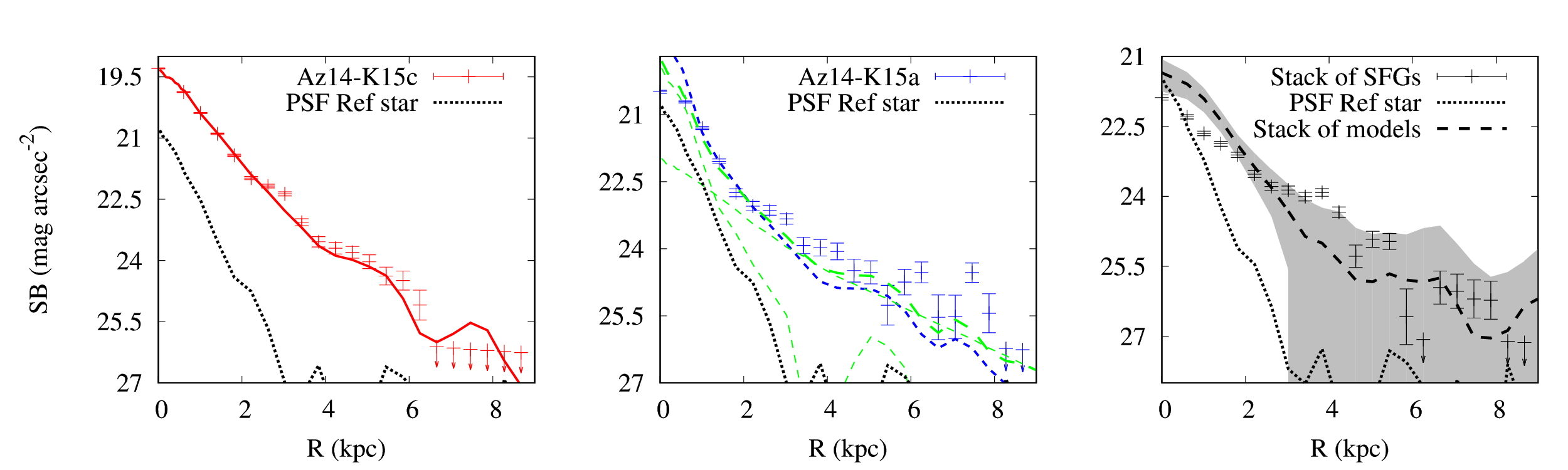}
\caption{
{\it Left:} The red crosses and solid line show the observed 
radial surface brightness profile 
and its best-fit model of Az14-K15c. 
The black dot line shows the radial profile of the PSF reference star. 
{\it Center:} 
The blue crosses and thick dashed line show the observed radial surface brightness profile 
and its best-fit model by a single S\'ersic model fitting of Az14-K15a.
The green thick long dashed line shows the best-fit model 
of the combination of a point source and S\'ersic model. 
The green thin dashed lines show a point source 
and S\'ersic model combined in the green thick long dashed line. 
{\it Right:} The black crosses show the radial surface brightness profile 
of a stack of the IRCS-AO $K'$-band images 
of the SFGs in the AzTEC14 group with 2$\sigma$ errors. 
The black dashed line and gray shaded region show 
the median and 2$\sigma$ rms around the median 
of the stacks of model galaxies with $r_e=0.5-2$ kpc and $n=0.5-4$, 
measured from a thousand times iteration. } 
\label{fig:rprofile}
\end{center} 
\end{figure*}

\begin{figure*}
\begin{center}
\includegraphics[width=140mm]{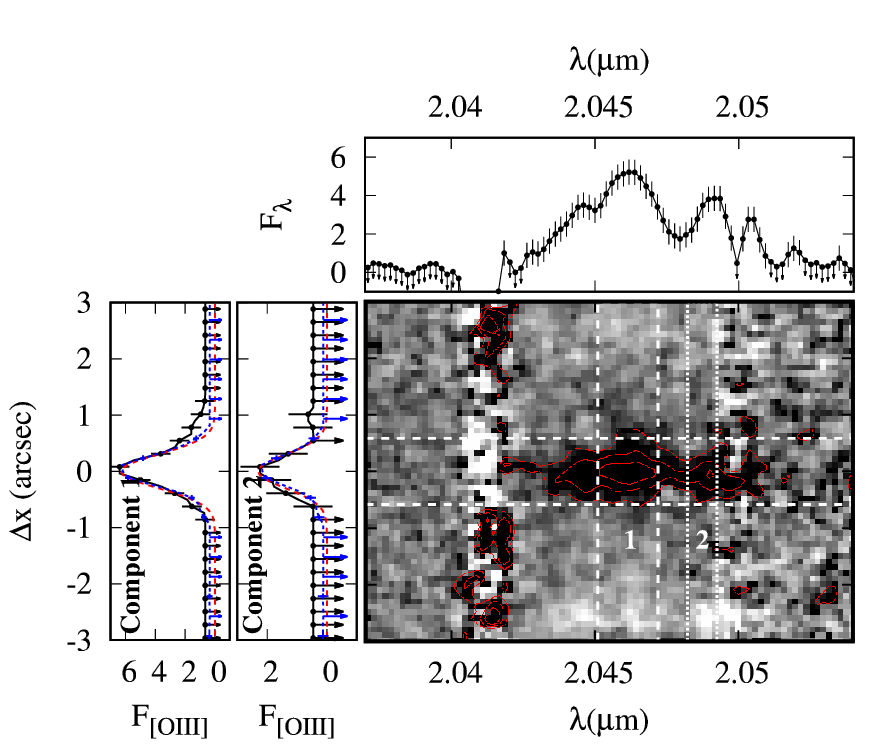}
\caption{
{\it Bottom right:} The spectral image of Az14-K15a around its [OIII]$\lambda5007$ emission lines. 
The red contours are isophotal area detected above 1.5, 3 and 5$\sigma$ per pixel. 
{\it Top: } The black points with lines show one dimensional spectrum  of Az14-K15a obtained
by summing up the spectrum image in spatial direction 
for $\approx 1''.2$ between the horizontal white dashed lines in the spectral image.
$F_{\lambda}$ is in $10^{-18}$ erg cm$^{-2}$ s$^{-1}$ \AA$^{-1}$.
{\it Bottom left:} The black points with lines show spatial extents 
of the spectrum at shorter (Component 1) and longer wavelength (Component 2) peaks
obtained by summing up the spectral image 
for $\approx 14~\&~7$ \AA~ in spectral direction, 
between the white vertical dashed lines and dot lines in the central image, respectively. 
$F_{\rm [OIII]}$ are in $10^{-18}$ erg cm$^{-2}$ s$^{-1}$pix$^{-1}$ (1pix=$0''.117$).
Component 2 is stacked avoiding adjacent OH sky emission.
The blue dot lines and cross points show the spatial extent 
of Az14-K15a measured on the MOIRCS $K_s$-band image 
smoothed to have FWHM PSF size $0''.7$, typical FWHM PSF size of this spectroscopic observation. 
The fluxes within the slit expected from our observing plan 
of this spectroscopic observation, where slit width was $0''.8$,  
are summed up and scaled to match with the scale of spectra. 
The red dashed lines show a spatial extent of a point source with FWHM PSF size $0''.7$, 
similarly measured as the blue dot lines and crosses.
} 
\label{fig:3dplot}
\end{center} 
\end{figure*}

The S\'ersic indices and effective radii of the QGs, Az14-K15a and -K15c 
are $n=9.5\pm4.5$ and $r_e=1.37\pm0.75$ kpc, 
and $n=2.5\pm0.2$ and $r_e=1.01\pm0.04$ kpc, respectively.
Fig. \ref{fig:residual} shows the observed, model and residual images 
of Az14-K15a and Az14-K15c obtained by using the {\sf GALFIT}. 
We show radial profiles of Az14-K15a and -K15c in the {\it central} 
and {\it left} panels of Fig. \ref{fig:rprofile}. 
Az14-K15c is well characterized as a massive compact elliptical similar to QGs found at $z>1$ 
(e.g., \citealt{2005ApJ...626..680D}; \citealt{2006ApJ...650...18T, 2007MNRAS.382..109T}; \citealt{2007ApJ...671..285T}; \citealt{2008ApJ...677L...5V, 2010ApJ...709.1018V}; \citealt{2009ApJ...695..101D}) but there is an matter of concern for Az14-K15a; it has an X-ray detected AGN detected with {\it Chandra} \citep{2009MNRAS.400..299L} which can sharpen its radial profile. 

Since it is hard to resolve an AGN from a high redshift galaxy spatially with current instruments, 
here we deal with the AGN component of Az14-K15a assuming that AGN to stellar flux ratio 
is comparable to emission line to continuum flux ratio at $K$-band. 
Fig. \ref{fig:3dplot} shows a spectrum and spatial extent 
of the [O{\footnotesize III}]$\lambda5007$ 
of Az14-K15a obtained in \citet{2015ApJ...799...38K}. 
It has double peaks in spectral direction but is just an point like source in spatial direction. 
The [O{\footnotesize III}]$\lambda5007$ line width and fluxes  
of the shorter and longer wavelength peaks are
$\sigma=225\pm11$ km s$^{-1}$ and $19.1\pm0.7\times10^{-17}$ erg cm$^{-2}$ s$^{-1}$, 
and $102\pm17$ km s$^{-1}$ and $6.5\pm0.8\times10^{-17}$ erg cm$^{-2}$ s$^{-1}$, respectively.
Together with H$\beta$ and [O{\footnotesize III}]$\lambda4959$, 
emission line flux contribute to about 30\% of the $K$-band flux. 
Then we re-fit the Az14-K15a to models composed from a point source
and a S\'ersic model with a fixed flux ratio of $3:7$, 
assuming that all the emission line flux shifted in the $K$-band 
is originated in AGN(s) (central point source) 
and all the continuum emission comes from the stellar component (S\'ersic model). 
Note that the influence of such a nuclear component is negligible for Az14-K15c 
since no signature of an AGN is detected 
and upper limit of the contamination of the [O{\footnotesize III}] 
to the $K$-band flux is less than 1\% in Az14-K15c 
from our spectroscopic observations in \citet{2015ApJ...799...38K}. 

The S\'ersic indices and effective radii of the best-fit S\'ersic model of double-components fit 
are $n=2.2\pm0.6$ and $r_e=4.3\pm0.6$ kpc. 
The green thin and thick long dashed lines in the {\it central} panel of Fig. \ref{fig:rprofile} show the best-fit 
models of a point source and S\'ersic model, and the sum of the two models. 
The model and residual images of the single and double components fits 
are shown in the {\it second} and {\it third} panels of Fig. \ref{fig:residual}. 
Both fits looks very similar but at a large scale, 
observed radial profile is reproduced better by double-components model. 

From these above, we found that one of the two QGs 
in the AzTEC14 group is as compact as field one at $z\sim3$ 
while other one is as large as giant elliptical at $z=0$, 
though there is large uncertainty in the latter one. 
It is reported that QGs in protoclusters
have sizes larger than those in general fields at $z<2$, 
implying accelerated size growths of them in overdense regions, 
possibly by enhanced merger rates 
\citep{2012ApJ...744..181Z, 2012MNRAS.419.3018C, 2012ApJ...750...93P, 2013ApJ...773..154L,2014ApJ...788...51N, 2016A&A...593A...2A}. 
At first, our results challenge the reliability of past studies arguing 
compact morphologies of QGs at high redshift without NIR spectroscopic follow-ups.  
In case of the SSA22 protocluster, 
now five, about a third of the candidate QGs selected in \citet{2013ApJ...778..170K}
are spectroscopically confirmed as the protocluster members in \citet{2015ApJ...799...38K}. 
Except for Az14-K15c, they are X-ray detected and confirmed by detecting 
their redshifted [O{\footnotesize III}]$\lambda5007$ 
or Ly$\alpha$ emission lines plausibly originated in AGNs, 
similar to Az14-K15a.
Other studies also report that AGNs are frequently seen 
among massive QGs at $z>2$ \citep{2013ApJ...764....4O,2016arXiv160605350M}. 
On the other hand, some massive compact ellipticals at high redshift are certainly found 
like Az14-K15c and in other studies by deep spectroscopic observations
(e.g., \citealt{2008ApJ...677L...5V}). 
Further studies of morphologies properly dealing with AGN components 
are required to conclude when giant ellipticals appeared in cluster of galaxies. 

It should be noted that [O{\footnotesize III}]$\lambda5007$ emission line of Az14-K15a 
at shorter wavelength has a wing at the upper side with respect to the centre (Fig. \ref{fig:3dplot}).
Since no such component is detected on both the $K$-band images, 
this wing component should have a large [O{\footnotesize III}]$\lambda5007$ 
equivalent width like Ly$\alpha$ Blobs. 
This wing component extends to $\sim1''\sim8$ kpc, much more larger than 
the typical size ($r_e=1\sim2$ kpc) of galaxies at $z\sim3$. 
We will discuss the origin of this extended and double peaked 
[O{\footnotesize III}]$\lambda5007$ emission lines in \S~4. 2. 

\subsubsection[Star forming galaxies]{Star forming galaxies}
The SFGs except for Az14-K15b and ADF22.4 are shown 
with black filled squares in Fig. \ref{fig:az14-re} 
while Az14-K15b is too diffuse to obtain a reasonable fit 
and ADF22.4 is not significantly detected on the IRCS-AO $K'$-band image. 
The stack of the SFGs is shown with the black open square. 
Except for the lowest stellar mass one, 
they tend to be larger than normal SFGs at the same redshift. 
In addition, the region detected over 2$\sigma$ per pixel of 
Az14-K15b ($M_{\star}\approx10^{11}~M_{\odot}$) 
is extended to $1''.5$ ($\approx11.5$ kpc) at least. 
Although these SFGs are too faint ($K=23.1\sim24.1$) 
to constrain their morphological parameters robustly, 
it is interesting that massive SFGs are all above 
the size-stellar mass relation of field galaxies. 
The simple photometric analysis in Fig. \ref{fig:moircs-ao2} 
also supports this tendency. 

The {\it right} panel of Fig. \ref{fig:rprofile} shows 
the radial profile of stack of the SFGs in the AzTEC14 group 
compared with the stacks of model galaxies in a range of typical galaxies at $z\sim3$.
We simulate stacked images by stacking five model galaxies 
with magnitude distributions same as that of the SFGs in the AzTEC14 group 
and morphological parameters randomly scattered in ranges of 
$r_e=0.5-2$ kpc, $n=0.5-4$, axis ratio $0\sim1$ and position angle $0\sim90$. 
The black dashed line and gray shaded region show the median 
and 2$\sigma$ rms around the median of radial profiles of the stacks of model galaxies. 
The radial profile of stack of the SFGs in the AzTEC14 group is
flatter than those of the model stacks of typical $z\sim3$ galaxies,  
i.e., the observed radial profile cannot be reproduced 
unless they are dominated by galaxies with $r_e>2$ kpc.
From these above, we conclude that massive SFGs in the AzTEC14 group
have the sizes on average larger than those of typical SFGs at $z\sim3$. 

The difference in the size-stellar mass relation between SFGs in the AzTEC14 group
and in general fields is likely to be originated in the sample bias  
since SFGs in the AzTEC14 group are classified as DRGs, 
namely rare massive dusty starburst galaxies. 
The size-stellar mass distribution of our sample 
is similar to that of massive H$\alpha$ emitters (HAEs) at $z=2.5$ \citep{2014ApJ...780...77T}.
In their study, low mass HAEs are on the size-stellar mass relation 
of normal SFGs at similar redshift while massive HAEs 
are on the size-stellar mass relation of SFGs at $z=0$. 
They searched HAEs at $z=2.5$ in general fields by using a narrow-band filter
but the strong spatial clustering of HAEs implies
that they are plausible progenitors of massive ETGs in clusters or groups,  
similar to the SFGs in the AzTEC14 group. 

The environmental dependence of morphologies of LBGs 
is discussed in many studies while the galaxies studied here 
are biased to DRGs, which do not often overlap with LBGs \citep{2013ApJ...778..170K}. 
We found no significant difference between C50, 
classified as a LBG, and field galaxies at $z\sim3$, 
similar to \citet{2007ApJ...668...23P} and \citet{2008ApJ...673..143O} 
while \citet{2016MNRAS.455.2363H} reported the enhanced 
merger fraction among the LBGs in the SSA22 protocluster.   
These studies used the images at rest-frame UV. 
Further observations at rest-frame optical may be also needed 
to discuss the environmental dependence of galaxy morphologies. 
On the other hand, \citet{2007ApJ...668...23P} reported 
that DRGs have relatively wide range of morphological parameters 
and include more high multiplicity objects and compact objects compared to LBGs. 
The environmental dependence of galaxy morphologies is still controversial, 
but at least, we can argue that DRGs, strongly clustered massive galaxies 
or plausible progenitors of massive ETGs today, 
have morphologies different from LBGs. 

\subsubsection[Sub-mm sources]{Sub-mm sources}

\citet{2015ApJ...810..133I} reported a median circularized sizes 
of $r_e\sim0.67$ kpc for bright SMGs measured at NIR, 
which is comparable to the effective radii of compact ellipticals at $z=2-3$. 
On the other hand, \citep{2016arXiv160707710R} reported that 
relatively faint ($S_{\rm 1~mm} \la1$ mJy) 
sub-mm sources at $z=2-3$ have the sizes of $r_e\sim4-5$ kpc at sub-mm 
in median, comparable to those of their stellar contents. 

There are five moderately luminous ($S_{\rm 1~mm} \sim1$ mJy) 
sub-mm sources identified by using ALMA at the AzTEC14 group 
and three of them are spectroscopically confirmed at $z_{\rm spec}\approx3.09$. 
We show the stamps of four out of the five sub-mm sources 
(K15b$=$ADF22.16, K15e$=$ADF22.11, ADF22.4 and ADF22.10) 
in Fig. \ref{fig:az14ao-image} while no significant counterparts 
are detected for the rest one, ADF22.17, in all the $I_{\rm F814W}$ and $K$-bands. 
The green contours in Fig. \ref{fig:az14ao-image} show the isophotal 
contours of the 1.1 mm sources detected above 3 $\sigma$ per beam. 
The blue solid lines of Az14-K15b in Fig. \ref{fig:az14ao-image} 
show the slit with $0''.8$ width used in our MOIRCS spectroscopy. 
As the alignment accuracy of MOIRCS is better than 0.1 arcsec 
rms\footnote{http://subarutelescope.org/Observing/Instruments/MOIRCS\\/spec\_mos.html},  
we might confirm the stellar component at the north 
of the sub-mm source and/or the sub-mm source itself. 
The bright object detected near ADF22.4 is a galaxy confirmed 
at $z_{\rm spec}\approx0.57$ \citep{2015ApJ...799...38K} 
and ADF22.4 can be lensed by this object. 

Although the $K'$-band counterparts of the sub-mm sources are 
too faint to constrain the robust morphological parameters with the {\sf GALFIT}, 
there are several clews suggesting that they have spatially extended stellar components; 
Az14-K15b is extended to $1''.5$ ($\approx11.5$ kpc). 
The $\approx85\%$ of the total flux of Az14-K15e 
is lost on our IRCS-AO $K'$-band image  
and possible counterparts of ADF22.4 and ADF22.10 
are detected on our MOIRCS $K_s$-band image
but not detected on our IRCS-AO $K'$-band image. 
The influence of morphologies on detection completeness 
and measured total flux values on our IRCS-AO $K'$-band image are described in Appendix C. 
Large sizes of them can cause such poor detections on our IRCS-AO $K'$-band image.  
In \citet{2017ApJ...835...98U}, it is reported that 
the deconvolved angular size of ADF22.4 is $1.9\pm0.4$ kpc in FWHM 
and those of other sub-mm sources in the SSA22 protocluster are $2-3$ kpc 
though the spatial resolution and signal to noise ratio are not enough 
to show whether they have remarkably compact or not.
At this point, it is not also clear whether there is a segregation 
between the morphologies in sub-mm and rest-frame optical. 

\subsection[Deficiency of low mass galaxies]{Deficiency of low mass galaxies}

Stellar mass function is one of the key properties to characterize a group of galaxies. 
Deep $K$-band images are useful to constrain stellar mass function of galaxies at $z>3$. 
In our previous study with MOIRCS \citep{2016MNRAS.455.3333K}, 
we show that stellar mass function of the AzTEC14 group
is consistent with those of proto-BCG groups at $z\sim3$ predicted 
from the galaxy formation models based on the Millennium simulation
\citep{2005Natur.435..629S,2007MNRAS.375....2D, 2011MNRAS.413..101G} 
at above $M_{\star}\geq4\times10^{10}~M_{\odot}$. 

Given the empirical size-stellar mass relation at $z\sim3$ in general fields, 
galaxies with stellar masses below the above completeness limit 
have sizes smaller than 1 kpc ($\approx0''.1$) in typical. 
If so, our new IRCS-AO $K'$-band image could give further constraints 
on stellar mass function of the AzTEC14 group; 
the completeness limit downs to $M_{\star}\ga 10^{10}~M_{\odot}$ 
and the $20\sim40$\% of galaxies with $M_{\star}=10^{9.5}\sim 10^{10}~M_{\odot}$ 
are expected to be detectable on our IRCS-AO $K'$-band image. 
Then $5\sim20$ new members should be detected 
if stellar mass function of the AzTEC14 group 
is continuously consistent with the above cosmological numerical simulations.  
However no additional member is detected in our IRCS-AO $K'$-band image. 

Detection completeness of galaxies depends on both colors and morphologies. 
The influence of colors as red as $J-K\sim2$ is already included 
in the above stellar mass completeness limit. 
Given that many of galaxies in the AzTEC14 group have sizes 
larger than typical $z\sim3$ galaxies, 
we cannot ignore the influence of source morphologies on detection completeness. 
MOIRCS $K_s$-band image may be also affected by source morphologies 
while our previous study ignored such an effect. 
We discuss stellar mass function of the AzTEC14 group in \S~4. 5.

\subsection[Fitting errors and detection completeness]{Fitting errors and detection completeness}

In this section, we test the influence of PSF variation, 
reproducibility of morphological parameters with the {\sf GALFIT}
and detection completeness on our IRCS-AO $K'$-band and MOIRCS $K_s$-band images. 

\subsubsection[PSF variation]{PSF variation}
Since performance of an AO system depends on separations 
of targets from a LGS and TTGS, 
we need to concern the influence of PSF variation 
on estimating morphological properties.
The separation between the LGS 
and the PSF reference star is 9 arcsec 
while those between the LGS 
and our targets range from $2-19$ arcsec.
According to the performance of the AO188\footnote{http://www.naoj.org/Observing/Instruments/AO/performance.html}, 
the FWHM of the PSF size at Az14-K15d can be $\sim10\%$ ($\sim0''.02$) 
smaller than that at the position of the PSF reference star 
while those at Az14-K15a, K15c and K15e can be $\sim5\%$ ($\sim0''.01$) smaller 
and that at Az14-K15f can be 10\% ($\approx0''.02$) larger.
The degradation of the performance owing to the separation 
from the TTGS is $\sim1\%$ for our targets.

To see the influence of PSF variation, 
we compare results of the {\sf GALFIT} obtained by using different PSF reference stars. 
Besides the PSF reference star adopted in this study, 
three stars are observed simultaneously 
but the FWHMs of the PSF sizes measured from them are $0''.16,~0''.17~\&~0''.18$. 
The separations of these stars from TTGS and LGS range $44-75$ and $18-23$ arcsec, respectively.
We compare the S\'ersic indices and effective radii 
estimated with these stars and the PSF reference star adopted 
in this study in Fig. \ref{fig:az14-testpsf}. 
The influence of PSF variation on the estimated effective radii is small 
except for a galaxy with the largest effective radius among the group. 
On the other hand, the estimates of S\'ersic indices vary greatly by PSF reference stars used. 

Again, at Az14-K15c, the brightest and one of the key objects of this study, 
the PSF size is expected to be $\sim5\%$ ($\sim0''.01$) 
smaller than that at the PSF reference star adopted. 
In Fig. \ref{fig:az14-testpsf}, use of $\sim0''.01$ different FWHM PSF size  
changes the estimated S\'ersic index and effective radius 
of Az14-K15c by only 0.5 and 0.2 kpc. 
Thus the results of Az14-K15c is not likely to be strongly affected 
by PSF variation, though ideally, we should also test of PSF references 
with PSF sizes smaller than that of the PSF reference star adopted in this study. 

\subsubsection[Performance of the {\sf GALFIT}]{Performance of the {\sf GALFIT} on our IRCS-AO $K'$-band image}

Next, we test the performance of the {\sf GALFIT} on our IRCS-AO $K'$-band image. 
To test the reproducibility of morphological parameters,
we generate mock galaxy images by making model galaxy 
images convolved with the observed PSF profiles 
by using the {\sf GALFIT} and putting them 
on the blank fields of the observed image to add the sky fluctuation. 
Then we re-run the {\sf GALFIT}. 
We show the deviations of measured values from initial values in Fig. \ref{fig:mkubo-az14-stdev-model}. 
Briefly, for typical compact elliptical galaxies at $z\sim3$ 
($n\geq 2.5$ and $r_e=0.5\sim1$ kpc) with $K_{\rm tot}=22$ ($K_{\rm tot}=22.5$), 
the 1$\sigma$ rms errors of the re-estimated values from the model parameters 
are $\sigma_{r_e} = 0.3~(0.5)$ kpc and $\sigma_n=1~(3)$.
For typical late-type galaxies at $z\sim3$ ($r_e=0.5\sim2$ kpc and $n<2.5$) 
with $K_{\rm tot}=22.5$ (24.0), 
the 1$\sigma$ rms errors are $\sigma_{r_e} = 0.7~(1.3)$ kpc and $\sigma_n=0.4~(1.9)$. 
For large late-type galaxies with $r_e=3$ kpc and $K_{\rm tot}=22.5~(24.0)$,  
the 1$\sigma$ rms errors are $\sigma_{r_e} = 1.2~(1.9)$ kpc and $\sigma_n=0.4~(2.8)$.
For faint objects with $K>22.5$ and $r_e>4$ kpc, measured $r_e$ are significantly underestimated. 
S\'ersic indices of objects with $K>22.5$ and/or large S\'ersic indices suffer from large errors.  

According to Table \ref{tab:tableobjects1}, and Fig. \ref{fig:mkubo-az14-stdev-model} and \ref{fig:mkubo-az14-model}, 
we can obtain reliable morphological parameters 
for Az14-K15c with $K_{\rm tot}=21.6$ 
but other members may suffer from large errors.
Az14-K15a is as bright as $K=22.5$ but 
as we describe above, its $K$-band flux is pushed up by an AGN component. 
$r_e$ of Az14-K15d, $n$ of Az14-K15f could be trusted  
though both are not reliable for Az14-K15e. 
There are possible additional uncertainties originated in the substructures 
like AGNs and giant clumps frequently seen among massive galaxies at high redshift 
(e.g., \citealt{2006ApJ...650..644E, 2006Natur.442..786G}, \citeyear{2008ApJ...687...59G}; 
S15; \citealt{2016ApJ...821...72S}). 
S15 reported that the fraction of clumpy galaxies increases with redshift 
and becomes $\sim50$\% at $z\sim2$. 
Multiple giant clumps are not clearly identified in our targets, 
may be due to the low signal to noise (S/N) ratio compared to previous studies with the {\it HST} 
but the different morphologies in different wavelength 
of Az14-K15b and C50 imply their complex structures. 

\subsubsection[Detection completeness]{Detection completeness}

Here we test dependence of detection completeness on source morphologies  
by generating mock galaxy images by the way described in \S~3.3.2
and extracting them by using the {\sf SExtractor}. 
We extract sources detected over $1.5~\sigma$ at each pixel 
($1~$pix$ =0''.052$ for IRCS and $0''.117$ for MOIRCS) for $>0.02$ and $>0.2$ arcsec$^2$ adjacent areas 
on the IRCS-AO $K'$ and MOIRCS $K_s$-band images, respectively. 

We show the detection completeness for models with 
$n=1~\&~4.0$ and $r_e=1,~2,~\&~3$ kpc 
in Fig. \ref{fig:mkubo-az14-detcomp-ao} \& \ref{fig:mkubo-az14-detcomp-mo}. 
Fig. \ref{fig:mkubo-az14-fluxcomp} shows the completeness of measured total magnitudes. 
The detection completeness on our IRCS-AO $K'$-band image 
sharply drops as sources have large sizes. 
Objects with $K_{\rm tot}<24$ are almost completely detectable on our IRCS-AO $K'$-band image 
but their total magnitudes can be significantly underestimated.  
The detection completeness on our MOIRCS $K_s$-band image 
is less sharply but also affected by source morphologies. 

\section[Discussion]{Discussion}
\subsection[Dense group of galaxies at high redshift]{Dense group of galaxies at high redshift}
There are many dense groups of massive 
galaxies in the SSA22 protocluster,  
which are plausible evidences of formations  
of giant elliptical galaxies via hierarchical multiple mergers. 
Such galaxy groups are reported in other studies. 
A dense cluster at $z=2.506$ (CL J1001) found by \citet{2016ApJ...828...56W} 
is similar to the AzTEC14 group in many aspects. 
The CL J1001 cluster has a collapsed halo with $M_{200c}=10^{13.9\pm0.2}~M_{\odot}$ 
and contains 11 massive galaxies within 80 kpc from the cluster centre 
($M_{\star}\ga10^{11}~M_{\odot}$ with the Salpeter IMF, 
corresponds to $M_{\star}\ga10^{10.5}~M_{\odot}$ with the Chabrier IMF). 
They found no object comparable to the CL J1001 cluster in the Millennium simulation 
while the AzTEC14 group has only one comparable group at each $z\sim3$ snapshot. 
The CL J1001 cluster and the AzTEC14 group are also similar in overdensities of DRGs, 
QGs, sub-mm sources and AGNs. 

QGs and many SFGs in the CL J1001 cluster 
are compact while no massive compact SFG is seen in the AzTEC14 group. 
Several studies reported compact SFGs and sub-mm galaxies at high redshift
(e.g., \citealt{2015ApJ...799...81S, 2015ApJ...810..133I,2015ApJ...811L...3T})
which can evolve into compact QGs by just quenching their star formation activities. 
The absence of compact SFGs in the AzTEC14 group 
can be owing to the short periods of compact SFG phases. 
We note that larger $k$-correction can be required in \citet{2016ApJ...828...56W} 
since they applied the morphological $k$-correction based on vdW14  
but their sample is biased to red galaxies rarely seen in general fields 
like galaxies in the AzTEC14 group which show very different morphologies 
in rest-frame UV and optical (Fig. \ref{fig:az14ao-image}) 
and careful subtractions of nuclei components with strong [O{\footnotesize II}] emission lines
can be required since many galaxies in the CL J1001 cluster show signatures of AGNs. 

It is beyond the scope of this paper but it is surprising that such rare density peaks 
hardly seen in the volume of the current large cosmological numerical simulations 
are discovered by field surveys with limited volumes. 
Further large volume cosmological numerical simulations 
and wide field surveys of such dense groups are required 
to show the consistency of their simulated and observed properties. 

\subsection[Extended and double peaked {[OIII]}$\lambda$5007 emission lines from a quiescent galaxy]{Extended and double peaked {[OIII]}$\lambda$5007 emission lines from a quiescent galaxy}

Since Az14-K15a is a young QG, to be strict, post-starburst galaxy
having the SED of $\sim0.5$ Gyr after burst-like star formation, 
footprints of its quenching process can be still observed. 
It is interesting that Az14-K15a is an AGN showing 
double peaked and spatially extended [O{\footnotesize III}]$\lambda$5007 emission lines. 

Plausible origins of double peaked emission lines from an AGN(s)
are dual AGNs, outflows and/or a rotating narrow line region
(e.g., \citealt{2011ApJ...739...69M, 2015ApJ...813..103M}). 
Since the two peaks have different line width, a rotating narrow line region scenario is ruled out. 
It is hard to resolve dual AGNs at $z>3$ with current instruments 
but if double peaked [O{\footnotesize III}]$\lambda$5007 emission lines 
are originated in dual AGNs, it is a direct evidence of major mergers of galaxies 
(or giant stellar clumps) hosting a SMBH for each. 
Similarly, the objects with double peaked CO emission lines are 
seen in a protocluster at $z=2.5$ \citep{2014ApJ...788L..23T} 
and a dense compact cluster of \citet{2016ApJ...828...56W}, 
suggesting frequent gas rich mergers of galaxies in the protocluster environments. 

An spatially extended metal emission line region 
is more likely to be produced by outflows rather than inflows of pristine gas.
Thus the [O{\footnotesize III}]$\lambda$5007 
of Az14-K15a is likely to be originated in outflows 
or a combination of outflows and dual AGNs.
It is very interesting to find a plausible signature of outflows 
from an AGN(s) in a QG in a protocluster at $z>3$. 
Similarly, [O{\footnotesize III}] ([O{\footnotesize II}]) Blobs were found at $z=1\sim1.6$
\citep{2013ApJ...765L...2B, 2013ApJ...779...53Y,2014ApJ...794..129H}. 
Further deep spatially resolved spectroscopy of Az14-K15a and 
other post-starburst galaxies in protoclusters 
may help us understanding how AGN activities relate with quenching of galaxies. 

\subsection[Evolution scenario of a BCG]{Evolution scenario of a BCG}

Discovery of a massive compact elliptical, Az14-K15c,  in such a proto-BCG group 
supports the two-phase formation scenario of giant elliptical galaxies 
that massive compact ellipticals formed at once and  
they evolve in sizes and stellar masses by series of mergers (e.g., \citealt{2010ApJ...725.2312O}), 
which has also been supported by many observational studies 
(e.g., \citealt{2010ApJ...709.1018V, 2015ApJ...805...34M, 2012ApJ...744..181Z, 2012MNRAS.419.3018C, 2012ApJ...750...93P, 2013ApJ...773..154L,2014ApJ...788...51N, 2016A&A...593A...2A}). 

It is known that BCGs are on the size-stellar mass relation above that of non-BCGs
\citep{2007AJ....133.1741B, 2009MNRAS.395.1491B,2015MNRAS.453.4444Z}. 
Several simulations (e.g., \citealt{2013MNRAS.435..901L};\citealt{2015ApJ...802...73S}) 
and observations (e.g., \citealt{2013MNRAS.433..825L,2013MNRAS.434.2856B,2015MNRAS.453.4444Z}) 
argue that BCGs double their stellar masses between $z=0$ to $1$ 
while little growths of BCGs at $z<2$ 
are reported in e.g., \citet{2009Natur.458..603C}, \citet{2010ApJ...718...23S, 2011MNRAS.414..445S}. 
Recently, \citet{2016ApJ...816...98Z} reported 
a redshift-dependent BCG-cluster mass relation at up to $z\sim1.2$. 
Az14-K15c needs to evolve in a size and stellar mass for $>10$ and four times, respectively 
to evolve into a BCG hosted in one of the most massive clusters 
today ($M_{\rm cluster}\sim10^{15}~M_{\odot}$ \& 
$M_{\star, \rm BCG}\sim10^{12}~M_{\odot}$)
while size growths by factor $\sim5$ from $z=0$ to $3$ 
are expected for compact QGs at $z>2$ in typical 
(e.g., \citealt{2007ApJ...671..285T, 2008ApJ...677L...5V, 2009ApJ...695..101D, 2010ApJ...709.1018V, 2014ApJ...788...28V}). 

We roughly estimate the size and stellar mass growths of Az14-K15c
assuming that all the group members will merge into this object. 
The size growth of an object by mergers 
can be written as $r_{g,f} / r_{g,i} = (1+\eta)^2/(1+\eta \epsilon)$,
adopting the virial theorem following \citet{2009ApJ...699L.178N}. 
$r_{g,f}$ and $r_{g,i}$ are the final and initial gravitational radii.   
$\eta$ and $\epsilon$ are ratios of masses and mean square speeds of the stars, respectively, 
between accreting and initial objects. 
Here we assume that the velocity dispersion of each group member 
is similar to those of compact QGs at $z=2-3$ 
which are 200 to 500 km s$^{-1}$ for galaxies 
with the stellar masses $M_{\star}=10^{10}-2\times10^{11}~M_{\odot}$, 
extrapolated from \citet{2009Natur.460..717V} and \citet{2009ApJ...697.1290B}. 
The above formalism is hold in case of dry mergers. 
Note that if massive SFGs in the group are gas rich when they merge, 
the size growth of Az14-K15c by mergers can be suppressed \citep{2015arXiv150205053W}. 

All the galaxies in the AzTEC14 group can merge into one massive galaxy in $3-4$ Gyr or by $z\sim1$, 
according to the numerical simulations of compact groups 
(e.g., \citealt{1989Natur.338..123B, 1993ApJ...416...17B}) 
which have the dynamical timescales similar to the AzTEC14 group. 
If we just use the observed stellar mass values of the members at $z=3.09$, 
the size and stellar mass of the final product 
are the $3-4$ times and double of the initial values of Az14-K15c, 
respectively (Case A in Fig. \ref{fig:az14-re}).  
It is consistent with the simulations and observations predicting continuing strong size growths at $z<1$. 
If the SFGs in the AzTEC14 group keep to be on or 
above the star formation main sequence for $\ga1$ Gyr before mergers, 
they can double their stellar masses and exhaust large fractions of gas.
In such case, the size and stellar mass of the final product 
are $8-10$ and four times of the initial values, respectively (Case B in Fig. \ref{fig:az14-re}). 
If Az14-K15c follows case B scenario and/or there are large extra accretion of galaxies, 
it could have already become a BCG today by $z>1$.  

Contributions of mergers of satellites to the evolution of a compact elliptical from $z=2$ to $1$
was discussed by using a deep {\it HST} image in \citet{2016ApJ...816...87M}. 
They argue that to reproduce the observed and simulated size growths of massive ETGs, 
not only just merging satellites but also in situ star formation in them are required. 
In case B, the net stellar mass increase by mergers 
is $\sim20\%$ of the descendant galaxy at $z\sim1$, 
similar to the results of \citet{2016ApJ...816...87M}.
Note that further size and stellar mass growths by satellites are expected for Az14-K15c 
since some of the members of the AzTEC14 group are sub-mm galaxies 
which may more rapidly grow in stellar masses than galaxies on the star formation main sequence,
much more accretions of satellites are expected at the core of a protocluster 
and the stellar mass completeness limit of our observations is not as small as \citet{2016ApJ...816...87M}. 

We note that some mergers expected in the AzTEC14 group 
can have the stellar mass ratios categorized as major mergers. 
Multiple major mergers can form slow rotators frequently seen 
among the most massive ETGs like BCGs 
while minor and major binary mergers result in fast rotators \citep{2014MNRAS.444.1475M}. 

\subsection[Nascent red sequence galaxies]{Nascent red sequence galaxies}

Not only compact QGs but also SFGs in protoclusters
are plausible progenitors of massive ETGs today. 
According to the model predictions, most of the members of the AzTEC14 group 
plausibly merge into one BCG in the current Universe 
but they still inform us how stars formed in galaxies at early time. 
Especially, massive SFGs in the AzTEC14 group are mostly classified as DRGs, 
known to show strong clustering \citep{2007ApJ...654..138Q,2007PASJ...59.1081I}, 
i.e., preferentially inhabiting in the environments where evolve into clusters or groups. 

One plausible explanation for the large sizes of massive SFGs classified as DRGs 
is the differences in halo masses 
before their dark matter halos incorporate into one massive halo. 
The sizes and rotational velocities of galaxy discs follow 
the sizes and circular velocities of their host dark matter halos.
It is known that DRGs show strong clustering \citep{2007ApJ...654..138Q,2007PASJ...59.1081I}. 
More strongly clustered galaxies are hosted in more massive dark matter halos. 
Based on the clustering analysis, \citet{2007ApJ...654..138Q} reported 
the halo mass $M_{\rm H}\sim5\times10^{12}~M_{\odot}$ 
and $M_{\rm H}\sim2\times10^{13}~M_{\odot}$ for photo-z or spec-z selected galaxies 
and DRGs with $K<21$ (in Vega, $\approx22.8$  in AB) at $2<z<3.5$, respectively. 
According to \citet{2013ApJ...770...57B}, 
the mean stellar to halo mass ratio peaks at $M_{\rm H}\sim10^{12}~M_{\odot}$ 
where $M_{\star}/M_{\rm H}$ is $\sim0.02$ and $M_{\star}$ is $\sim2\times10^{10}~M_{\odot}$. 
When the mass of a halo increases from $M_{\rm H} =10^{12}$ 
to $10^{13}~M_{\odot}$, the stellar mass inside increases by only three times. 
Thus there is no wonder that halo masses of galaxies 
with stellar masses $M_{\star}=10^{10.5}-10^{11}~M_{\odot}$ range widely. 
Both the size and circular velocity of a halo
are proportional to a cubic root of the halo mass, 
approximated with the spherical collapse model (Eq. (2) of \citealt{1998MNRAS.295..319M}). 
Thus DRGs would have the disc sizes (and rotational velocities) 
twice larger than those of normal SFGs,  
consistent with the observed size difference of massive SFGs in the AzTEC14 group. 

The sizes and stellar masses of massive SFGs in the AzTEC14 group 
are comparable to those of massive ETGs in the local Universe (Fig. \ref{fig:az14-re}). 
It is interesting to find a post starburst galaxy, Az14-K15a, 
composed of an AGN component and flat stellar component. 
The size and stellar mass growths via mergers can be less important 
for them to evolve into local massive ETGs. 
Even though they have late-type morphologies at this point, 
they can evolve into fast-rotating ETGs by exhausting gas (e.g., \citealt{2011MNRAS.417..845K}).
On the other hand, remarkable compactness of Az14-K15c implies 
the two-phase formation scenario in which massive compact ellipticals are formed at once 
and evolve into local massive ETGs through many mergers in later \citep{2010ApJ...725.2312O}.  
Gas rich major mergers 
\citep{2006ApJ...650..791C,2007ApJ...658..710N, 2010ApJ...722.1666W,2011ApJ...730....4B}, 
and inflowing gas and wet mergers of inflowing clumps 
via disc instabilities \citep{2009ApJ...703..785D, 2015MNRAS.447.3291C} 
can form massive compact ellipticals from large discy SFGs. 
Further studies of the relation between morphologies, 
stellar populations and also AGN activities in protoclusters, 
including such large and red SFGs, 
are needed to understand how massive SFGs transformed into massive ETGs today. 

\subsection[Stellar mass function and deficiency of low mass galaxies]{Stellar mass function and deficiency of low mass galaxies: The AzTEC14 group is a more massive group?}

Given that many galaxies in the AzTEC14 group have sizes $r_e>2$ kpc, 
source morphology is likely to affect severely on the detection completeness 
on not only our IRCS-AO $K'$-band but also MOIRCS $K_s$-band images. 
This suggests that stellar mass function obtained with MOIRCS in previous study 
is less complete than that we expected; 
stellar mass completeness limit can be twice larger ($\sim10^{11}~M_{\odot}$) 
and total flux values can be underestimated to reduce stellar masses estimates. 
Then the AzTEC14 group can be a group richer than that we claimed in previous study. 
Unfortunately, there may be no such a massive group  
in current large volume cosmological numerical simulations, 
since comparison groups of AzTEC14 group found in previous study 
is the richest groups in the Millennium simulation. 
Note that \citet{2009MNRAS.395..114H} reported that 
stellar mass function around the QSO at the center 
of a protocluster at $z=2.16$, also a plausible progenitor of a BCG,  
is consistent with the galaxy formation models 
based on the Millennium simulation at $M_{\star}>10^{10}~M_{\odot}$. 
But their results may be less affected by morphologies of galaxies 
since their targets are at the redshift lower than our target, 
i.e., less affected by cosmological surface brightness dimming, 
and they used the {\it HST} with the Strehl ratio higher than that of the AO188 in typical. 

Much more deep imaging observations are required to measure stellar mass function robustly 
and show whether the observed deficiency of faint galaxies 
is originated in diffuse morphologies and/or actual deficiency of them. 
Suppressions of formations of low-mass galaxies 
by reionization \citep{2000ApJ...539..517B} 
and supernovae feedbacks \citep{2002MNRAS.333..177B, 2003ApJ...599...38B}
are predicted but the strength of such effects are still open question. 
The large stellar mass existing in the group 
implies that it had the star formation density higher than that in general fields at past, 
i.e., there is a strong UV radiation field heating the gas of low mass halos.
But the characteristic halo mass at which a halo lost half of its baryon 
by the reionization or supernovae feedback 
is too low ($M_{\rm H}\sim 10^{9}~M_{\odot}$) to cause the deficiency 
of faint galaxies observed in the AzTEC14 group. 
On the other hand, supernovae feedbacks
can work effectively to transport angular momenta from inner 
to outer-radii of galaxies to extend their sizes \citep{2002MNRAS.333..156B}.
It can also happen that gravitational heating prevents the cooling of low mass sub-halos 
since the AzTEC14 group can be (partly) virialized 
as it has the halo mass $M_{\rm H}\sim10^{13}~M_{\odot}$ \citep{2016MNRAS.455.3333K}.
 
\section[Conclusion]{Conclusion}

We conducted the deep and high resolution imaging 
of an extremely dense group of galaxies, called the AzTEC14 group, 
at the core of the protocluster at $z=3.09$ in the SSA22 field 
by using the AO188/IRCS equipped on Subaru Telescope 
to study morphological evolution of massive ETGs. 
Wide morphological variety of them implies 
that morphology-density relation today has just begun forming. 

We confirm that one of the two QGs in the group, the most massive member, is a compact QG. 
This supports the two-phase formation scenario of giant elliptical galaxies 
that massive compact ellipticals formed at once and  
they evolve in sizes and stellar masses by series of mergers. 
To form a local BCG-like object by $z\sim1$, sometimes observed, 
in situ star formation in the group members may be important. 
Another QG in the group 
is fitted with the model composed of a nuclear component 
and not so compact S\'ersic model, 
and shows double peaked and spatially extended [O{\footnotesize III}]$\lambda$5007 emission lines ($\sim8$ kpc). 
It is a key result to find an evidence of outflows from an AGN(s) in a young QG.  
Massive SFGs in the group 
have stellar masses and sizes comparable to those of local massive ETGs. 
Even if massive SFGs become compact spheroids at once by gas rich major mergers, 
large stellar masses of them imply the importance 
of star formations before violent morphological evolutions. 
Although we obtained the image more sensitive for typical $z\sim3$ galaxies in general fields 
than our previous MOIRCS $K_s$-band image, no candidate new group members are detected. 
It implies that there is an actual deficiency of low mass galaxies 
and/or they are too diffuse to be detected on our IRCS-AO $K'$-band image. 
Moreover, given the morphological trend of the AzTEC14 group found in this study, 
our previous estimate of stellar mass function of the AzTEC14 group with MOIRCS 
is likely to be less complete than that we expected. 

We argue the necessity of more careful treatments of diffuse and red galaxies at $z>3$ 
which are hardly detected with the {\it HST} 
but may play important roles in massive galaxy formation. 
More deep and wide imaging surveys at wavelength longer than $2~\mu$m with large telescopes
are needed to study such red, diffuse and faint galaxies. 
Careful subtractions of AGNs from compact QGs and SFGs are also important 
to evaluate the size evolution history of massive ETGs correctly. 

\section*{Acknowledgments}

This study is based on data collected at Subaru Telescope, 
which is operated by the National Astronomical Observatory of Japan. 
We would like to thank the Subaru Telescope staff 
for many help and support for the observations. 
Our studies owe a lot deal to the archival Subaru 
Suprime-Cam (\citealt{2004AJ....128..569M}), {\it Spitzer} IRAC \& MIPS data taken in
\citet{2009ApJ...692.1561W}, {\it Chandra} data taken in
\citet{2009MNRAS.400..299L}. 
We also thank to AzTEC/ASTE observers of the SSA22 field 
providing the updated source catalog.  
This work was supported by Global COE Program "Weaving
Science Web beyond Particle-Matter Hierarchy", MEXT, Japan. 
YM acknowledges support from JSPS KAKENHI Grant Number 20647268. 
This work was partially supported by JSPS Grants-in-Aid for Scientific  Research No.26400217.
HU is supported by the ALMA Japan Research Grant of NAOJ Chile Observatory, 
NAOJ-ALMA-0071, 0131, 140, and 0152. 
HU is supported by JSPS Grant-in-Aid for Research Activity Start-up (16H06713). 
This paper makes use of the following ALMA data: ADS/JAO.ALMA\#2013.1.00162.S. ALMA is a partnership of ESO (representing its member states), NSF (USA) and NINS (Japan), together with NRC (Canada) and NSC and ASIAA (Taiwan) and KASI (Republic of Korea), in cooperation with the Republic of Chile. The Joint ALMA Observatory is operated by ESO, AUI/NRAO and NAOJ.

\appendix
\section{Influence of PSF variation}
In this work, we perform two-dimensional fits of galaxies 
adopting an observed image of the closest star to the image center  
as a PSF reference. 
Since performance of an AO system is not uniform on the whole image, 
we need to concern influence of PSF variation on evaluating morphological properties.
According to the performance of the AO188, 
FWHM of PSF sizes at our targets can be $-0''.02$ to $+0''.02$ 
different from that at the star adopted as a PSF reference in this study. 

\begin{figure*} 
\begin{center}
\includegraphics[width=65mm]{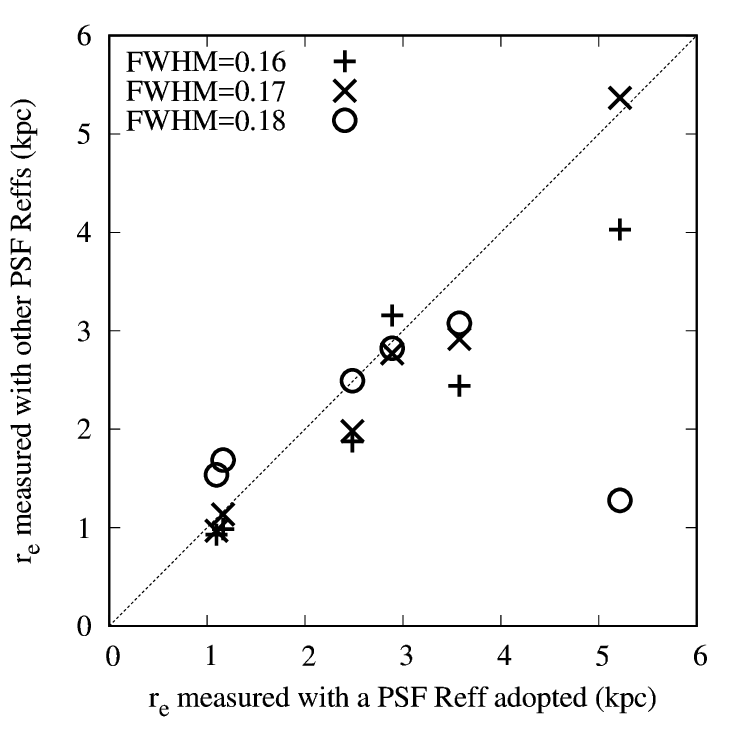}
\includegraphics[width=65mm]{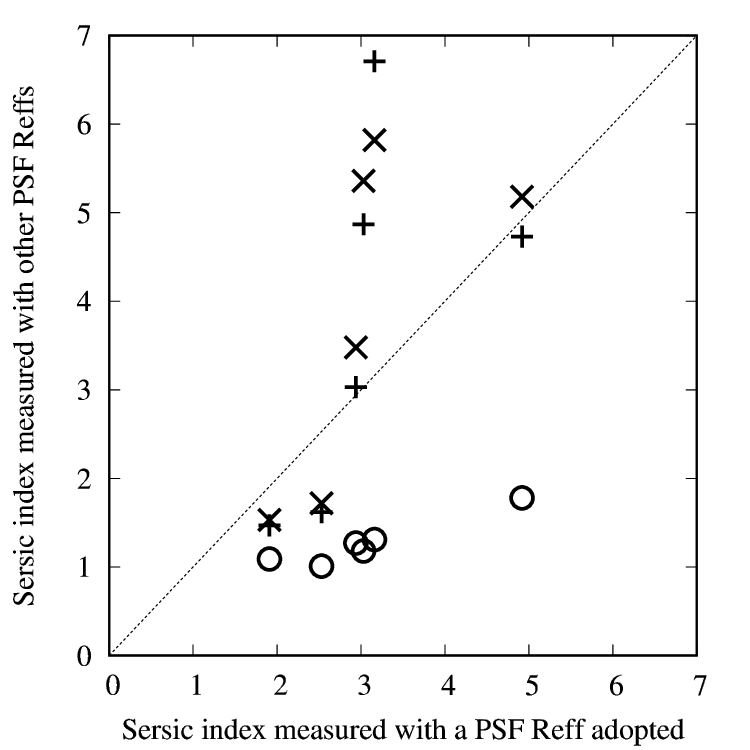}
\caption{
{\it Left:} Effective radii of galaxies estimated 
with the PSF reference star adopted in this study v.s. 
those estimated using other stars as PSF references. 
The crosses, x-marks and circles show 
in cases using stars with the FWHMs of the PSF sizes
$0''.16, 0''.17~\&~0''.18$, respectively, as PSF references. 
{\it Right:} Similar to the {\it left} panel but shows S\'ersic indices $n$.
} 
\label{fig:az14-testpsf}
\end{center}
\end{figure*}

To test influence of PSF variation, 
we compare two-dimensional fits performed 
with different stars found in our image as PSF references. 
(R.A., Dec), FWHM PSF sizes, separations from 
TTGS and LGS of these stars are as follows; 
(22:17:37.644 +0:18:06.71), $0.16$, 75 and 15 arcsec;
(22:17:36.457 +0:18:34.00), $0.17$, 44 and 18 arcsec;
(22:17:35.648 +0:18:24.29), $0.18$, 52 and 23 arcsec. 
Fig. \ref{fig:az14-testpsf} shows morphological properties 
evaluated by adopting different PSF references. 
Effective radii and S\'ersic indices of galaxies measured 
with a star other than the PSF reference star adopted in this study are shown 
against those estimated by using the PSF reference star adopted in this study. 

The influence of PSF variation on estimated effective radii is small 
except for a galaxy with the largest effective radius among the group. 
On the other hand, the estimates of S\'ersic indices vary greatly by selections of PSF references. 

\section{Reproducibility of morphological parameters}
Here we test reproducibility of morphological properties with the {\sf GALFIT}.
To test reproducibility of morphological properties,
we generate mock galaxy images by making model galaxy images 
convolved with the observed PSF profiles 
by using the {\sf GALFIT} and putting them on the blank fields 
of the observed image to add the sky fluctuation. 
Then we re-run the {\sf GALFIT}. 

\begin{figure*} 
\begin{center}
\includegraphics[width=58mm]{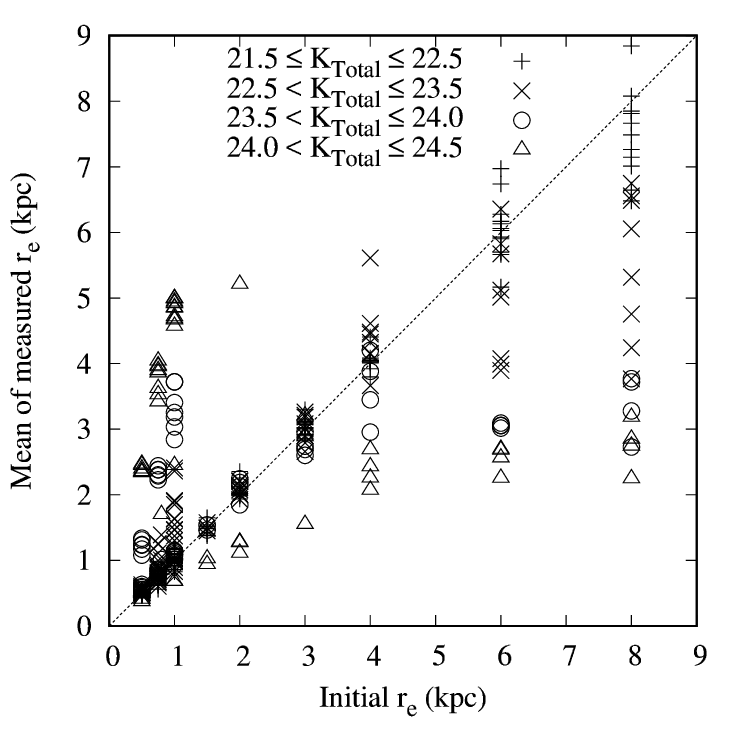}
\includegraphics[width=58mm]{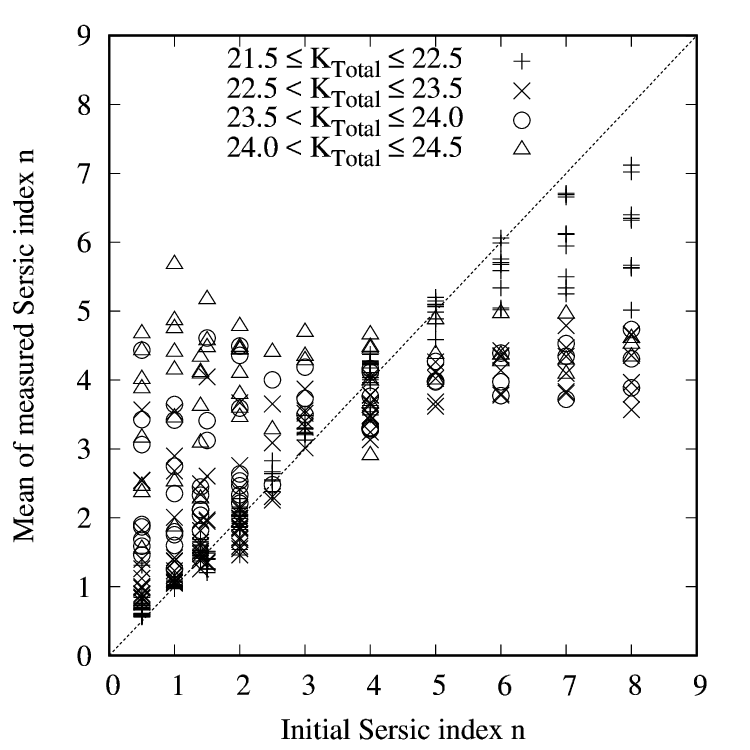}
\includegraphics[width=58mm]{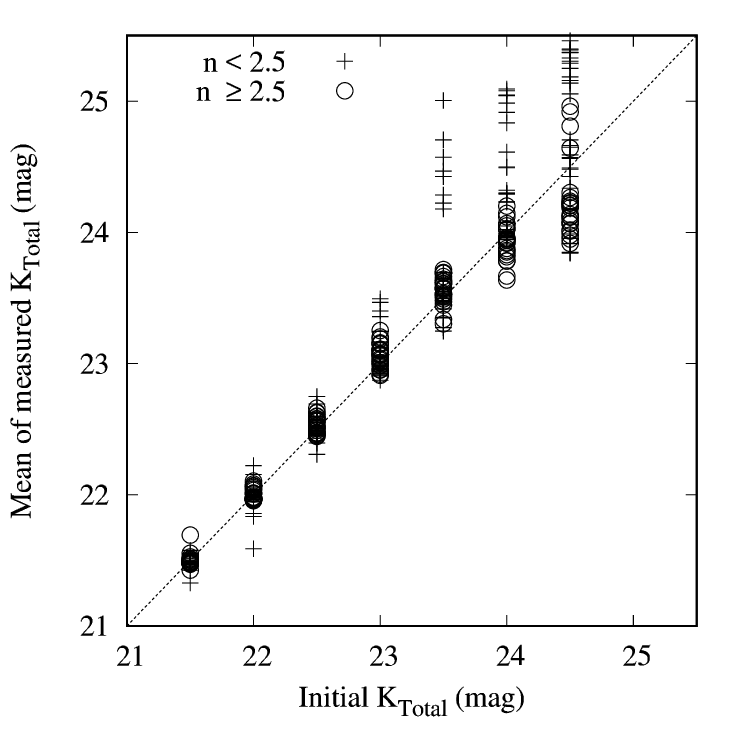}
\caption{
{\it Left:} Initial input effective radii v.s. means of effective radii 
measured on simulated images for models with $r_e=1-8$ kpc, 
$n=0-8$ and $21.5 \leq K_{\rm tot} \leq 24.5$.
The crosses, x-marks, circles and triangles show results for 
initial inputs of $21.5 \leq K_{\rm tot} \leq 22.5$,  $22.5 < K_{\rm tot} \leq 23.5$,  
$23.5 < K_{\rm tot} \leq 24.0$ and $24.0 < K_{\rm tot} \leq 24.5$, respectively. 
{\it Center:} Similar to the {\it left} panel but for S\'ersic indices. 
{\it Right:} Similar to the {\it left} panel but for total magnitudes. 
The crosses and circles show results for initial inputs of $n<2.5$ and $n>2.5$, respectively.
} 
\label{fig:mkubo-az14-model}
\end{center}
\end{figure*}

\begin{figure*} 
\begin{center}
\includegraphics[width=58mm]{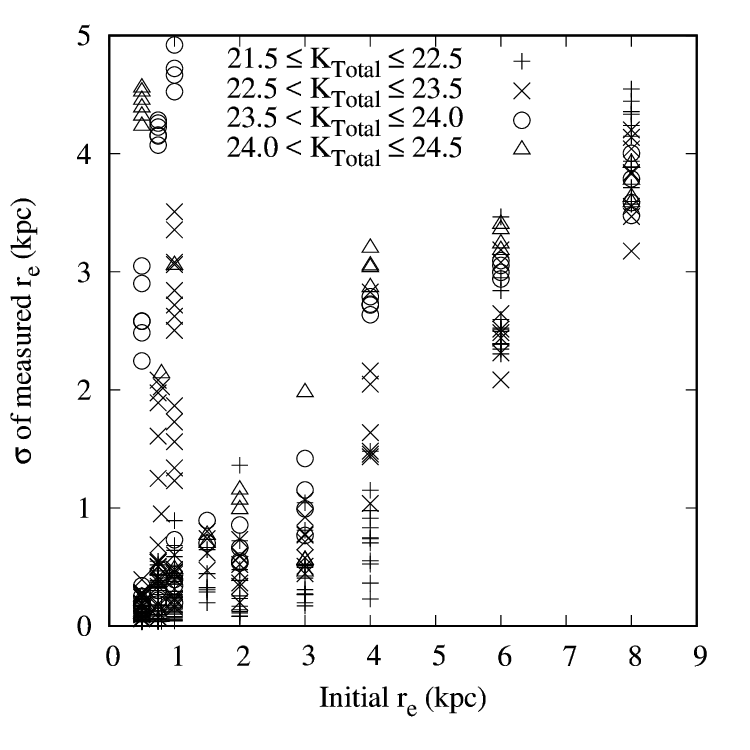}
\includegraphics[width=58mm]{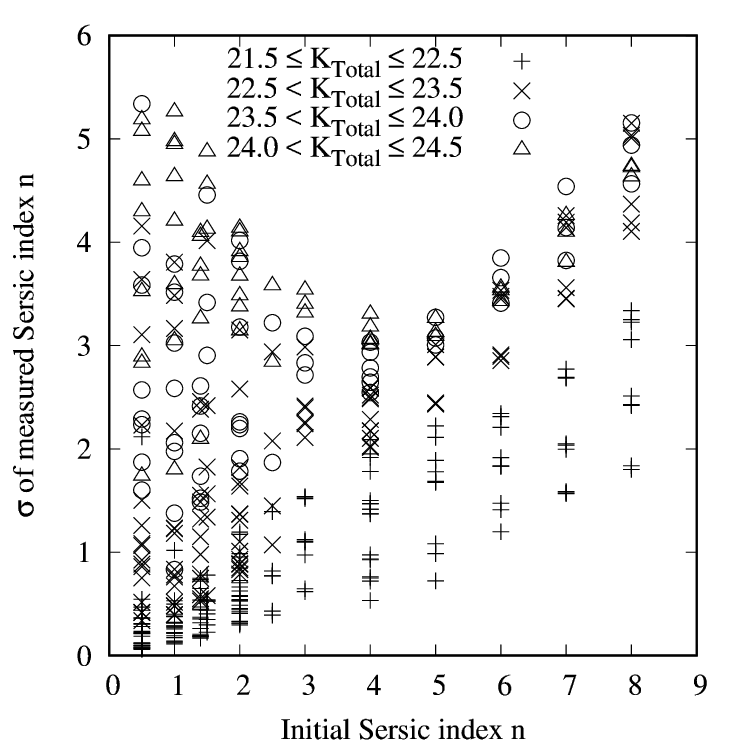}
\includegraphics[width=58mm]{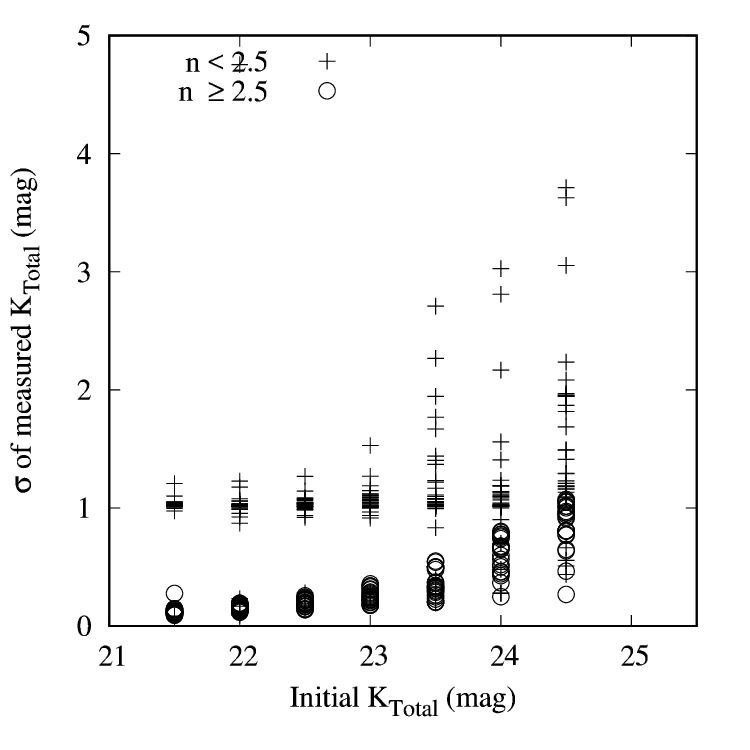}
\caption{
Similar to Fig. \ref{fig:mkubo-az14-model} 
but standard deviations of measured values from initial inputs are shown in y-axis.
} 
\label{fig:mkubo-az14-stdev-model}
\end{center}
\end{figure*}

We test the S\'ersic models with the S\'ersic indices ranging from $n=0.5$ to $8$, 
effective radii $r_e=0.5$ to $8$ kpc and total magnitudes $K_{\rm tot}=21.5-24.5$. 
We performed a thousand simulations for each model 
and see deviations of re-estimated values from inputs. 
Fig. \ref{fig:mkubo-az14-model} compares initial inputs and means 
of effective radii, S\'ersic indices and total magnitudes measured on simulated images. 
Fig. \ref{fig:mkubo-az14-stdev-model} compares initial inputs and standard deviation  
of effective radii, S\'ersic indices and total magnitudes measured on simulated images
from their initial inputs. 

Fitting errors get larger for models with fainter $K_{tot}$, larger $r_e$ and $n$.
For typical compact elliptical galaxies at $z\sim3$ 
($n\geq 2.5$ and $r_e=0.5\sim1$ kpc) with $K_{\rm tot}=22$ ($K_{\rm tot}=22.5$), 
the 1$\sigma$ rms errors of the re-estimated values from the model parameters 
are $\sigma_{r_e} = 0.3~(0.5)$ kpc and $\sigma_n=1~(3)$.
For typical late-type galaxies at $z\sim3$ ($r_e=0.5\sim2$ kpc and $n<2.5$) 
with $K_{\rm tot}=22.5$ (24.0), 
the 1$\sigma$ rms errors are $\sigma_{r_e} = 0.7~(1.3)$ kpc and $\sigma_n=0.4~(1.9)$. 
For large late-type galaxies with $r_e=3$ kpc and $K_{\rm tot}=22.5~(24.0)$,  
the 1$\sigma$ rms errors are $\sigma_{r_e} = 1.2~(1.9)$ kpc and $\sigma_n=0.4~(2.8)$. 

\section{Detection Completeness}

We test dependence of detection completeness on galaxy morphologies 
on both our IRCS-AO $K'$-band and MOIRCS $K_s$-band images
by generating mock galaxy images by the way described in Appendix B 
and extracting them by using the {\sf SExtractor} \citep{1996A&AS..117..393B}. 
For MOIRCS $K_s$-band image, the PSF convolved with model galaxies 
is generated from stars observed simultaneously with our targets, 
where FoV of single MOIRCS detector is $3'.5\times4'.0$,
by {\sf PSF} task of the {\sf IRAF}. 
We extract sources detected over $1.5~\sigma$ at each pixel 
($1~$pix $ =0''.052$ for IRCS and $0''.117$ for MOIRCS) for $>0.02$ and $>0.2$ arcsec$^2$ adjacent areas 
on our IRCS-AO $K'$ and MOIRCS $K_s$-band images, respectively. 

\begin{figure*} 
\begin{center}
\includegraphics[width=75mm]{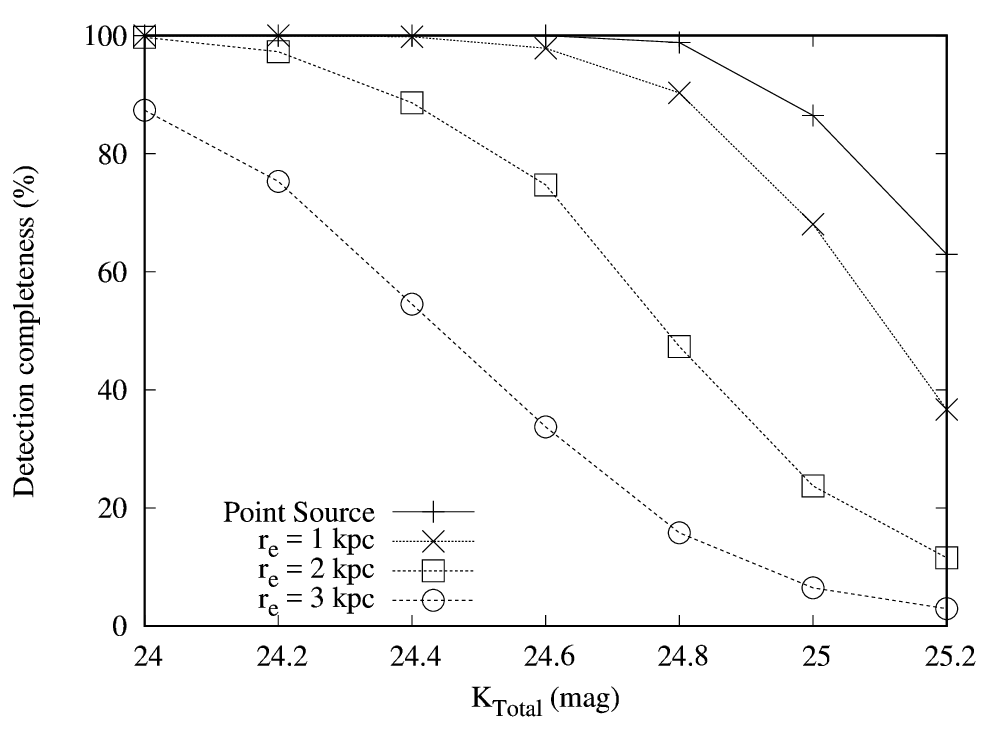}
\includegraphics[width=75mm]{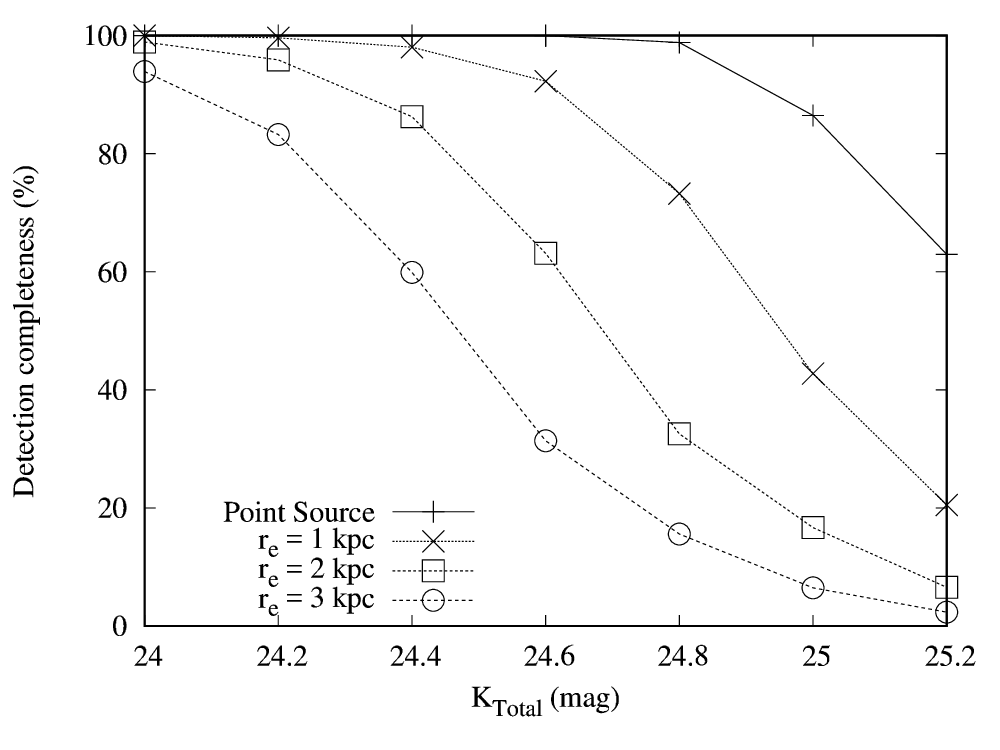}
\caption{
{\it Left:} Detection completenesses of models 
with S\'ersic indices $n=1$ on our IRCS-AO $K'$-band image. 
The x-marks, squares and circles with dot lines show detection completeness
of models with $r_e=1,~2~\&~3$ kpc, respectively. 
The cross-points with a solid line show that of point sources. 
{\it Right:} Similar to the {\it left} panel but for models with S\'ersic indices $n=4$. 
} 
\label{fig:mkubo-az14-detcomp-ao}
\end{center}
\end{figure*}

\begin{figure*} 
\begin{center}
\includegraphics[width=75mm]{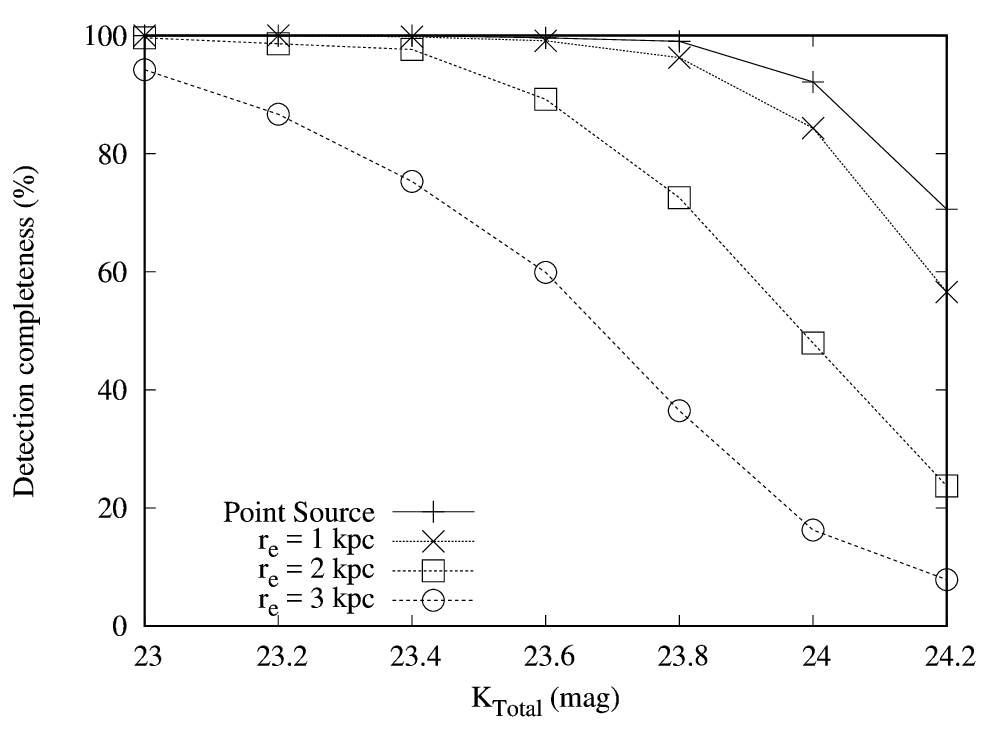}
\includegraphics[width=75mm]{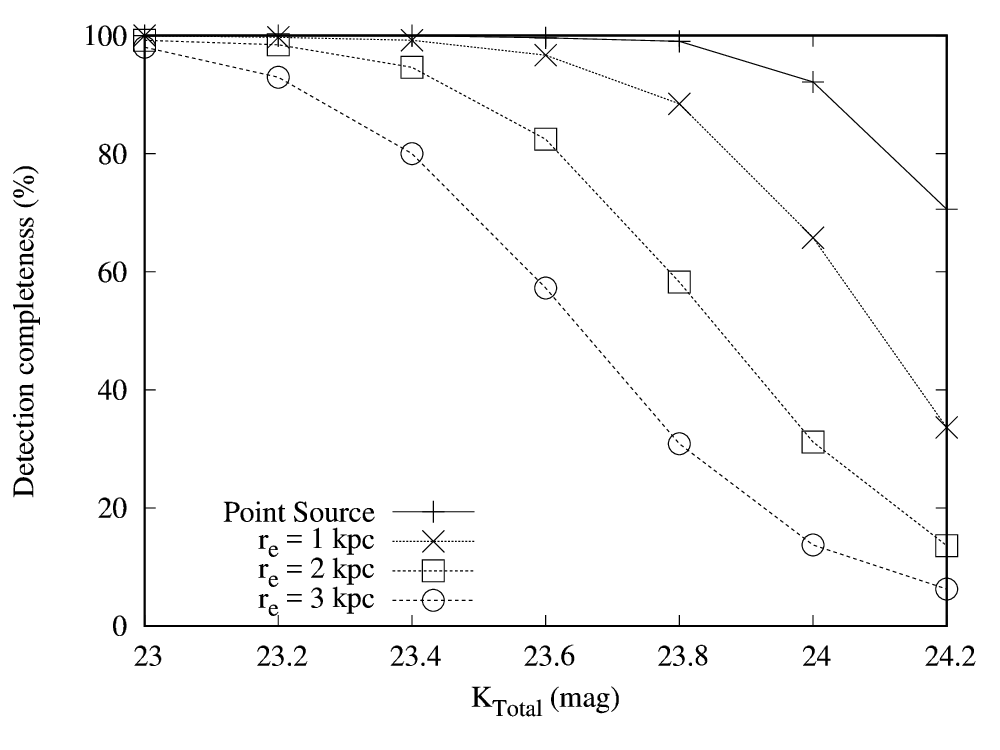}
\caption{
Similar to Fig. \ref{fig:mkubo-az14-detcomp-ao} but for our MOIRCS $K_s$-band image.
} 
\label{fig:mkubo-az14-detcomp-mo}
\end{center}
\end{figure*}

\begin{figure*} 
\begin{center}
\includegraphics[width=75mm]{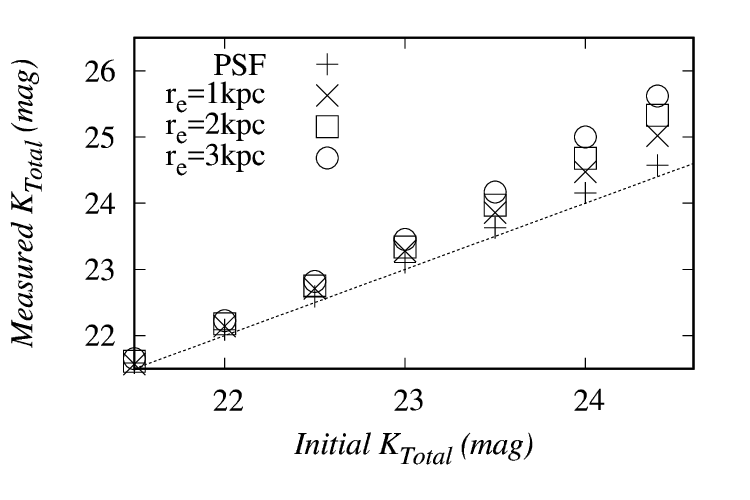}
\includegraphics[width=75mm]{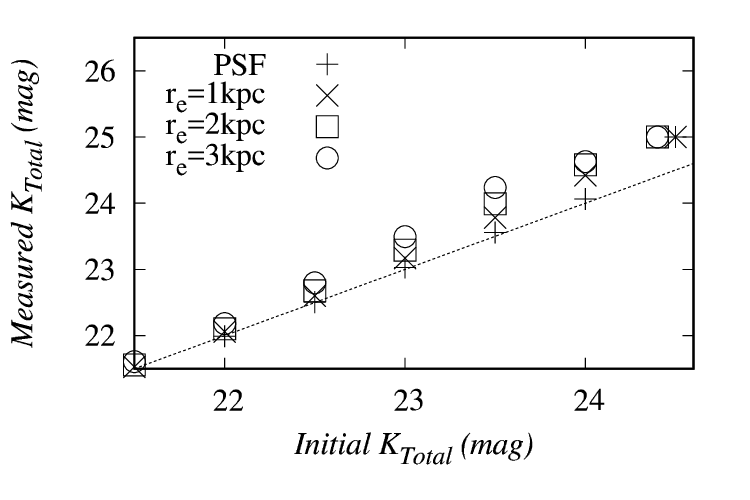}
\caption{
{\it Left:} Means of total magnitudes measured on simulated images of our IRCS-AO $K'$-band image
for models with S\'ersic indices $n=1$. 
The x-marks, squares and circles show those of models with $r_e=1,~2~\&~3$ kpc, respectively. 
The cross-points show those of point sources. 
Means are taken for objects successfully detected on simulated images.
The dot line shows the case that total magnitudes measured on simulated images
properly reproduce initial input values. 
{\it Right:} Similar to the {\it left} panel but for our MOIRCS $K_s$-band image. 
} 
\label{fig:mkubo-az14-fluxcomp}
\end{center}
\end{figure*}

Figure \ref{fig:mkubo-az14-detcomp-ao} and \ref{fig:mkubo-az14-detcomp-mo} 
show detection completeness  in cases $n=1~\&~4$ and $r_e=1\sim3$ kpc
on our IRCS-AO $K'$-band and MOIRCS $K_s$-band images, respectively. 
Fig. \ref{fig:mkubo-az14-fluxcomp} shows means of measured total magnitudes 
in cases $n=1$ and $r_e=1\sim3$ kpc on 
both IRCS-AO $K'$-band and MOIRCS $K_s$-band images. 

Impact of object morphologies on detection completeness 
is stronger for models with low S\'ersic indices on the IRCS-AO $K'$-band image. 
In addition, detection completeness declines as sizes of objects increase. 
There are also significant underestimation of total magnitudes 
depending on sizes. 
Object morphologies is less strongly but also affect 
detection completeness on our MOIRCS $K_s$-band image. 
Objects with $K_{\rm tot}<24$ are almost completely 
detectable on our IRCS-AO $K'$-band image 
but their total fluxes are likely to be significantly underestimated. 

\label{lastpage}

\end{document}